\documentclass[aps,prd,superscriptaddress,floatfix,nofootinbib,preprint,11pt]{revtex4-1}
\makeatletter
\def\l@subsubsection#1#2{}
\makeatother
\usepackage[english]{babel}
\usepackage[utf8]{inputenc}
\usepackage[T1]{fontenc}
\usepackage{amsmath,amssymb,braket,slashed,mathrsfs,mathtools}
\usepackage{pifont}

\usepackage{bm}

\usepackage[squaren,Gray,cdot,binary]{SIunits}
\usepackage{graphicx,subfigure}
\graphicspath{{./figures/}}
\usepackage{color,hyperref}
\usepackage{minibox}
\usepackage{cleveref}
\usepackage{marginnote}

\crefformat{section}{Sec.~#2#1#3}
\crefmultiformat{section}{Sec.~#2#1#3}
    { and~#2#1#3}{, #2#1#3}{ and~#2#1#3}
\crefrangeformat{section}{Sec.~#3#1#4\,--\,#5#2#6}
\crefformat{subsection}{Sec.~#2#1#3}
\crefmultiformat{subsection}{Sec.~#2#1#3}
    { and~#2#1#3}{, #2#1#3}{ and~#2#1#3}
\crefrangeformat{subsection}{Sec.~#3#1#4\,--\,#5#2#6}
\crefformat{subsubsection}{Sec.~#2#1#3}
\crefmultiformat{subsubsection}{Sec.~#2#1#3}
    { and~#2#1#3}{, #2#1#3}{ and~#2#1#3}
\crefrangeformat{subsubsection}{Sec.~#3#1#4\,--\,#5#2#6}
\crefformat{figure}{Fig.~#2#1#3}
\crefmultiformat{figure}{Figs.~#2#1#3}
    { and~#2#1#3}{, #2#1#3}{ and~#2#1#3}
\crefrangeformat{figure}{Figs.~#3#1#4\,--\,#5#2#6}
\crefformat{table}{Table~#2#1#3}
\crefmultiformat{table}{Tables~#2#1#3}
    { and~#2#1#3}{, #2#1#3}{ and~#2#1#3}
\crefrangeformat{table}{Tables~#3#1#4\,--\,#5#2#6}
\crefformat{equation}{Eq.~(#2#1#3)}
\crefmultiformat{equation}{Eqs.~(#2#1#3)}{ and~(#2#1#3)}{, (#2#1#3)}{ and~(#2#1#3)}
\crefrangeformat{equation}{Eqs.~(#3#1#4)--(#5#2#6)}
                
\definecolor{linkcolor}{rgb}{.17578125,.1875,.5703125}
\hypersetup{
    pdfstartview={FitH},                         
    pdftitle={Scalar EMT Renormalisation Using The Wilson Flow},                            
    pdfauthor={Joseph K. L. Lee},                         
    pdfsubject={Physics article},                              
    colorlinks=true,                             
    linkcolor=linkcolor,                         
    citecolor=linkcolor,                         
    filecolor=black,                             
    urlcolor=linkcolor      						 
}

\newcommand{\ie}{\textit{i.e.}~}

\let\ol\overline
\let\wh\widehat

\newcommand{\cB}{\mathcal{B}}
\newcommand{\ZZ}{\mathbb{Z}}

\DeclarePairedDelimiter\pq{\lparen}{\rparen}
\DeclarePairedDelimiter\bq{\lbrack}{\rbrack}

\newcommand{\pfrac}[2]{\pq*{\frac{#1}{#2}}}

\renewcommand{\d}{\mathrm{d}}
\newcommand{\dk}{\frac{\d^3 k}{(2\pi)^3}}

\newcommand{\ord}[1]{\mathcal{O}\left({#1}\right)}

\DeclareMathOperator{\Tr}{Tr}
\DeclareMathOperator{\erfi}{Erfi}
\DeclareMathOperator{\erf}{Erf}

\renewcommand{\bar}{\overline}

\newcommand{\diff}{\mathrm{d}}

\renewcommand{\epsilon}{\varepsilon}

\newcommand{\su}{\mathfrak{su}}

\DeclareMathOperator{\tr}{tr}

\newcommand{\dmom}[1]{\frac{\mathrm{d}^3{#1}}{(2\pi)^3}}

\newcommand{\suml}[1]{\sum_{#1\in\Lambda^3}}
\newcommand{\intlm}[1]{\int_{-\frac{\pi}{a}}^{\frac{\pi}{a}}
                       \frac{\diff^3#1}{(2\pi)^3}\,}

\newcommand{\overbar}[1]{\mkern 1.5mu\overline{\mkern-1.5mu#1\mkern-1.5mu}\mkern 1.5mu}
\newcommand{\mn}{{\mu\nu}}
\newcommand{\Tmn}{T_{\mu\nu}}
\newcommand{\TRmn}{T^R_{\mu\nu}}
\newcommand{\rs}{{\rho\sigma}}

\newcommand{\mnrs}{{\mu\nu\rho\sigma}}

\begin{document}
  \title{
  Renormalization of the energy-momentum tensor in \\
  three-dimensional scalar SU(N) theories using the Wilson flow
 }
\author{Luigi~Del~Debbio}
\affiliation{Higgs Centre for Theoretical Physics, School of Physics and Astronomy, The University of Edinburgh, Edinburgh EH9 3FD, United Kingdom}
\author{Elizabeth~Dobson}
\affiliation{Higgs Centre for Theoretical Physics, School of Physics and Astronomy, The University of Edinburgh, Edinburgh EH9 3FD, United Kingdom}
\affiliation{Institute of Physics, The University of Graz, Universit\"atsplatz 5, A-8010 Graz, Austria}
\author{Andreas~J\"uttner}
\affiliation{School of Physics and Astronomy, University of Southampton, Southampton SO17 1BJ, United Kingdom}
 \affiliation{STAG Research Center,
University of Southampton, Highfield, Southampton SO17 1BJ, United Kingdom}
\author{Ben~Kitching-Morley}
\affiliation{School of Physics and Astronomy, University of Southampton, Southampton SO17 1BJ, United Kingdom}
\affiliation{STAG Research Center, University of Southampton, Highfield, Southampton SO17 1BJ, United Kingdom}
\affiliation{Mathematical Sciences, University of Southampton, Highfield, Southampton SO17 1BJ, United Kingdom}
\author{Joseph~K.~L.~Lee}
\email{joseph.lee@ed.ac.uk}
\affiliation{Higgs Centre for Theoretical Physics, School of Physics and Astronomy, The University of Edinburgh, Edinburgh EH9 3FD, United Kingdom}
\author{Valentin~Nourry}
\affiliation{Higgs Centre for Theoretical Physics, School of Physics and Astronomy, The University of Edinburgh, Edinburgh EH9 3FD, United Kingdom}
\affiliation{STAG Research Center, University of Southampton, Highfield, Southampton SO17 1BJ, United Kingdom}
\affiliation{Universit\'e de Paris, CNRS, Astroparticule et Cosmologie, F-75006 Paris, France}
\author{Antonin~Portelli}%
\author{Henrique~Bergallo Rocha}
\affiliation{Higgs Centre for Theoretical Physics, School of Physics and Astronomy, The University of Edinburgh, Edinburgh EH9 3FD, United Kingdom}
\author{Kostas~Skenderis}
\affiliation{STAG Research Center, University of Southampton, Highfield, Southampton SO17 1BJ, United Kingdom}
 \affiliation{Mathematical Sciences,
University of Southampton, Highfield, Southampton SO17 1BJ, United Kingdom}

 \collaboration{LatCos Collaboration}
  \begin{abstract}
	A nonperturbative determination of the energy-momentum tensor is essential for understanding the physics of strongly coupled systems. The ability of the Wilson flow to eliminate divergent contact terms makes it a practical method for renormalizing the energy-momentum tensor on the lattice. In this paper, we utilize the Wilson flow to define a procedure to renormalize the energy-momentum tensor for a three-dimensional massless scalar field in the adjoint of $SU(N)$ with a $\varphi^4$ interaction on the lattice. In this theory the energy-momentum tensor can mix with $\varphi^2$ and we present numerical results for the mixing coefficient for the $N=2$ theory.
  \end{abstract}
  \maketitle
  \tableofcontents
  \section{Introduction}
  \label{sec:introduction}
  The energy-momentum tensor (EMT) plays a fundamental role in quantum field theories, by virtue of being the collection of Noether currents related to space-time symmetries. It acts as the source for space-time curvature in the Einstein field equations, and its expectation value encodes the energy and momentum carried by quantum excitations. One of the motivations for this study comes from the application of holography to cosmology~\citep{McFadden:2009fg}. In this holographic approach, cosmological observables, such as the cosmic microwave background (CMB) power spectra, can be described in terms of correlators of the EMT of a dual three-dimensional quantum field theory (QFT) with no gravity. The dual theories introduced in~\citep{McFadden:2009fg} comprise three-dimensional Yang-Mills theory, coupled to massless scalars $\varphi$ in the adjoint of $SU(N)$ with a $\varphi^4$ interaction. Perturbative calculations of the correlators have been performed~\citep{McFadden:2010na, McFadden:2010vh, Bzowski:2011ab,Coriano:2020zap} and the predictions of holographic cosmology were tested favorably against Planck data in~\citep{Afshordi:2016dvb}. The results in~\citep{Afshordi:2016dvb} however also implied that a nonperturbative evaluation of the EMT is required in order to fully exploit the duality in the low-multipole regime. 

Here we initiate the computation of nonperturbative effects by means of lattice QFT. A fundamental limitation of the lattice framework is the fact that space-time symmetries, such as Poincar\'{e} invariance, are explicitly broken at finite lattice spacing; these symmetries are restored only in the continuum limit. Consequently, the Ward identities associated with translations are violated, and the EMT, which generates such transformations, has to be defined with care. On the lattice, the EMT has to be renormalized by tuning the coefficients of a linear combination of all operators with dimension not greater than the space-time dimension $d$, which are compatible with lattice symmetries. This ensures that the Ward identities are recovered in the continuum limit, up to cutoff effects. Perturbative analytic calculations using this method have been discussed extensively in~\citep{Caracciolo:1988hc, Caracciolo:1989pt}.

Various strategies have been proposed to nonperturbatively renormalize the EMT on the lattice (cf.~\citep{Suzuki:2016ytc}, and references therein) such as the shifted boundary condition~\citep{Giusti:2012yj,Giusti:2010bb,Giusti:2015daa, DallaBrida:2020gux}, applying the Wilson flow on the EMT~\citep{Suzuki:2013gza,Asakawa:2013laa,Makino:2014taa,Makino:2014wca,Kitazawa:2014uxa,Kitazawa:2016dsl,Kitazawa:2017qab,Hirakida:2018uoy,Suzuki:2021tlr}, and on probe operators~\citep{Capponi:2015ucc,Capponi:2015ahp,Capponi:2016yjz}, which is the strategy considered in this paper. The Wilson flow~\citep{Narayanan:2006rf,Luscher:2009eq,Luscher:2010iy,Luscher:2011bx} has been used to renormalize composite operators in various scenarios~\citep{Luscher:2013vga,Ramos:2015dla,Monahan:2015lha,Fujikawa:2016qis,Carosso:2018bmz}. The method adopted here is to construct probes from fields at some positive flowtime, which are nonlocal in the elementary fields, that can eliminate the divergent contact terms present in the correlators. The divergence properties and regularization of Ward identities of flowed gauge fields are discussed extensively in~\citep{DelDebbio:2013zaa}.

In this paper we are interested in renormalizing the EMT of the simplest version of the holographic dual theories, which is the class of $3d$ massless scalar QFTs with $\varphi$ in the adjoint of $SU(N)$ and a $\varphi^4$ interaction, regularized on a Euclidean space-time lattice~\citep{Lee:2019zml}. This model is interesting in its own right. If correct, this model would provide a remarkably simple description of the very early Universe, with the microscopic theory containing only two parameters, $N$ and the nonminimality parameter $\xi$.\footnote{One should not confuse the number of parameters appearing in empirical models, such as the $\Lambda$CDM model with the number of parameters appearing in the microscopic theory. For example, $\Lambda$CDM contains two parameters associated with the very early Universe
(the amplitudes of primordial perturbations and the spectral index), but the underlying microscopic inflationary models contain a lot more parameters (the parameters appearing in the inflaton potential etc.)} Preliminary results show that it provides an excellent fit to CMB data in the regime where perturbation theory can be trusted, while suggesting that the model becomes nonperturbative at higher multipoles than the best fit model based on Yang-Mills theory coupled to scalars (roughly, $\ell \lesssim 250$ versus $\ell \lesssim 30$), so this model also serves as an example of a holographic model where the effective dimensionless coupling is of intermediate strength (neither very large nor very small) for a sizeable part of the CMB spectrum, and as such it requires a nonperturbative treatment. 
 
This class of massless, super-renormalizable QFT, with the coupling $g$ of mass dimension one, suffers from severe infrared (IR) divergences in perturbation theory. Perturbative calculations of correlation functions and renormalization parameters, such as the critical mass or the EMT renormalization coefficients, contain IR divergences, which makes the results dependent on the IR regulator. The nonperturbative IR finiteness of super-renormalizable theories, where the dimensionful coupling constant acts as the IR regulator, has been conjectured and discussed in~\citep{Jackiw:1980kv, Appelquist:1981vg}, and has been confirmed nonperturbatively for the theory under consideration in~\citep{Cossu:2020yeg}. This allows us to renormalize the theory nonperturbatively without IR ambiguity. The properties of $3d$ super-renormalizable scalar QFTs with various symmetry groups have been widely studied both perturbatively and on the lattice~\citep{Farakos:1994kx,Kajantie:1995dw,Arnold:2001ir,Sun:2002cc,Gynther:2007bw,Pelissetto:2015yha}. In this paper we focus on the $N=2$ theory; theories with $N>2$ and the large $N$ limit will be discussed in a later publication.

This paper is organized as follows. In~\cref{sec:generalities} we first introduce the scalar $SU(N)$ theory in the continuum and on the lattice, and we define the EMT operator and correlators. We also define the Wilson flow, as well as the relevant correlators at finite flowtime. In~\cref{sec:simulations} we list the parameters of the simulated ensembles for this study, and summarize the results of the critical mass determined nonperturbatively in~\citep{Cossu:2020yeg}. In~\cref{sec:renormalisation} we discuss the procedure to renormalize the EMT using flowed correlators, and finally present the numerical results for the $N=2$ theory. We have also included a number of appendixes. In~\cref{sec:app_eval_ints} we summarize the method to evaluate massless lattice scalar integrals in $3d$. In~\cref{sec:app_emt_c3,sec:app_c2cmn_0t,sec:app_c2cmn_ft}, we present the lattice perturbation theory calculations for the EMT operator mixing, correlators at vanishing flowtime, and correlators at finite flowtime respectively.

  \section{Generalities/Definitions}
  \label{sec:generalities}
    \subsection{Continuum and lattice $SU(N)$ scalar action}

	The theory under consideration here is a three-dimensional Euclidean scalar $\su (N)$ valued $\varphi^4$ theory,
	\begin{align}\label{eq:canonicalS}
		S\left[\varphi \right] =\int d^3x \Tr \left[  \left(\partial_\mu\varphi(x)\right)^2 + (m^2-m_c^2) \varphi(x)^2 + \lambda \varphi(x)^4 \right], 
	\end{align}
  with fields $\varphi = \varphi^a(x) T^a$ where $\varphi^a(x)$ is real, and $T^a$ are the generators of $SU(N)$, which are normalized so that $\Tr \left[ T^a T^b\right] = \frac{1}{2}\delta_{ab}$. Here $\lambda$ is the $\varphi^4$ coupling constant with mass dimension one (which does not renormalize), $m^2$ is the bare mass. Since the mass of the theory renormalizes additively, we include the mass counterterm, or \textit{critical mass} $m^2_c(g)$, i.e. the value of the bare mass such that the renormalized theory is massless. To make the 't Hooft scaling explicit, hereafter the following rescaled version of the action will be used,
  	\begin{align}
		S\left[ \phi \right] =\frac{N}{g} \int d^3x \Tr \left[  \left(\partial_\mu\phi(x)\right)^2 + (m^2-m_c^2) \phi(x)^2 + \phi(x)^4 \right], 
	\end{align}
	which can be obtained by identifying $\phi=\sqrt{g/N} \varphi$ and $\lambda = g/N$ from~\cref{eq:canonicalS}.

	The theory is discretized on a three-dimensional Euclidean lattice by replacing the action with
	 \begin{align}
		S\left[ \phi \right] =\frac{a^3N}{g} \sum_{x \in \Lambda^3} \Tr \left[  \sum_\mu \left(\delta_\mu\phi(x)\right)^2 + (m^2-m_c^2) \phi(x)^2 + \phi(x)^4 \right]. 
	 \end{align}
	 Here $\delta_\mu$ is the forward finite difference operator defined by, $\delta_\mu \phi(x) = a^{-1} \left[ \phi(x+a\hat{\mu}) - \phi(x)\right]$, where $\hat{\mu}$ is the unit vector in direction $\mu$, $\Lambda^3$ is a lattice with cubic geometry containing $N_L^3$ points (with periodic boundary conditions), and $a$ the lattice spacing.  

  \subsection{Energy-momentum tensor}

	In the continuum theory, the energy-momentum tensor $\Tmn$ is defined as the conserved current of space-time symmetries. For our scalar $SU(N)$ theory, it is given by~\citep{Collins:1976va}
	\begin{align}
		\Tmn = \frac{N}{g} \Tr \left\{ 2(\partial_\mu \phi)(\partial_\nu \phi) - \delta_\mn \left[ \sum_\rho (\partial_\rho \phi)^2 + (m^2 - m_c^2) \phi^2 + \phi^4 \right] + \xi \left(\delta_\mn \sum_\rho (\partial_\rho \phi)^2  - (\partial_\mu \phi)(\partial_\nu \phi)\right)\right\}.
\end{align}
Here the term multiplying $\xi$ is the improvement term. In the continuum theory, due to translational invariance, the EMT satisfies Ward-Takahashi identities (WI) of the form 
  \begin{align} \label{eq:continuum-WI}
  	\langle \partial^\mu \Tmn(x) P(y)\rangle = -\bigg \langle \frac{\delta P(y)}{\delta\phi(x)} \partial_\nu \phi(x) \bigg\rangle
  \end{align}
  where $P(y)$ is any composite operator inserted at point $y$. If $P$ is such that the right-hand side of~\cref{eq:continuum-WI} is finite for separated points  $x \neq y$, the left-hand side correlation function, which contains the divergence of the EMT, is finite up to contact terms. For this theory, it can be shown that the insertion of $\Tmn$ does not introduce new UV divergences (as discussed in more detail in~\cref{sec:app_emt_c3}). The improvement term is identically conserved and trivially satisfies~\cref{eq:continuum-WI}. Therefore $\xi$ will be set to 0 for the remainder of the text.

  On the lattice, the continuous translational symmetry is broken into the discrete subgroup of lattice translations; because of this a na\"ive discretization of the EMT on the lattice,
  \begin{align}
  	\Tmn^0 = \frac{N}{g} \Tr \left\{ 2(\overbar{\delta}_\mu \phi)(\overbar{\delta}_\nu \phi) - \delta_\mn \left[ \sum_\rho (\overbar{\delta}_\rho \phi)^2 + (m^2 - m_c^2) \phi^2 + \phi^4 \right] \right\},
  \end{align}
  which is obtained by replacing the partial derivatives $\partial_\mu \phi(x)$ with the central finite difference $\overbar{\delta}_\mu \phi(x) = \frac{1}{2a}\left[ \phi(x+a\hat{\mu}) - \phi(x-a\hat{\mu}) \right]$ (this is chosen in order to obtain a Hermitian EMT), does not satisfy the WI~\cref{eq:continuum-WI}. Now, the WI on the lattice includes an additional term~\citep{Caracciolo:1988hc},
    \begin{align} \label{eq:lattice-WI}
  	\langle \overbar{\delta}^\mu \Tmn^0 (x) P(y)\rangle = -\bigg \langle \frac{\overbar{\delta}P(y)}{\overbar{\delta}\phi(x)}\overbar{\delta}_\nu \phi(x) \bigg\rangle + \langle X_\nu (x) P(y) \rangle .
  \end{align}
  Here $\frac{\overbar{\delta}P(y)}{\overbar{\delta}\phi(x)}$ is obtained by replacing the fields and derivatives in the continuum functional derivative $\frac{\delta P(y)}{\delta\phi(x)}$ with their lattice counterparts, and $X_\nu$ is an operator proportional to $a^2$, which classically  vanishes in the continuum limit. However, radiative corrections cause the expectation value $\langle X_\nu (x) P(y) \rangle$ to produce a linearly $a^{-1}$ divergent contribution to the WI. Therefore, the na\"ively discretized EMT will not reproduce the continuum WI when the regulator is removed; $\Tmn$ has to be renormalized by adjusting the coefficients of a linear combination of lower-dimensional operators which satisfy the same symmetries.

  In four dimensions, it has been shown in~\citep{Caracciolo:1988hc} that $\Tmn$ potentially mixes with five lower-dimensional operators, which can generate such divergences. However, in three dimensions, dimensional counting indicates that divergent mixing can only occur with $O_3 = \delta_\mn \frac{N}{g}\Tr \phi^2$. The \textit{renormalized EMT} on the lattice can therefore be defined as an operator mixing,
  \begin{align} \label{eq:renormalised-emt}
  	\TRmn &= \Tmn^0 - C_3 \delta_\mn \frac{N}{g} \Tr \phi^2.
  \end{align}
$C_3$ has to be tuned to satisfy the continuum WI up to discretization effects when the regulator is removed.

At leading order (LO) $O(g)$ (i.e. one loop) in lattice perturbation theory, $C_3$ is shown to be
\begin{align} \label{eq:C3c3-1loop}
C_3^{\text{1 loop}} = \frac{g}{a}  c_3^{\text{1 loop}},
\end{align}
  where 
  \begin{align}\label{eq:c3-1loop}
c_3^{\text{1 loop}}&=\left(2-\frac{3}{N^2}\right)\left(\frac{6Z_0-1}{12}\right),  \\
  Z_0 &= a \intlm{k} \frac{1}{\hat{k}^2}  = 0.252731...,
\end{align}   
for lattice momentum $\hat{k} = \frac{2}{a} \sin (ka/2)$, see~\cref{sec:app_emt_c3}. In the continuum limit, $a \to 0$, the value of $C_3^{\text{1 loop}}$ diverges. To account for this leading behavior, we define
\begin{align} \label{eq:small-c3-def}
C_3 = \frac{g}{a}c_3,
\end{align}
and by determining the value of $c_3$ nonperturbatively, we are able to renormalize the EMT on the lattice. As mentioned in the Introduction, the two-loop contribution diverges logarithmically with the IR regulator.

	Before discussing the strategy to obtain the value of $c_3$ nonperturbatively, we define an EMT correlator which will be useful in our analysis. Consider the momentum-space two-point correlator,
	\begin{align} \label{eq:Cmn}
	C_\mn(q)  = \frac{N}{g} a^3 \sum_{x \in \Lambda} e^{-i q \cdot x} \langle \Tmn^R (x) \Tr \phi^2 (0) \rangle. 
\end{align}
Here $q=\frac{2\pi}{aN_L} n$ is the momentum where $n$ is a vector with integer components. This particular correlator is chosen since $\Tr \phi^2$ is the lowest dimension nonvanishing scalar operator in the theory. By inserting the definitions in~\cref{eq:renormalised-emt,eq:small-c3-def}, we obtain
\begin{align} \label{eq:Cmn-c3C2}
	C_\mn(q)  &= C_\mn^0(q) - \frac{g}{a}c_3 \delta_\mn  C_2(q),
\end{align} where
\begin{align}
C_\mn^0(q) = \frac{N}{g} a^3 \sum_{x \in \Lambda} e^{-i q \cdot x} \langle \Tmn^0 (x) \Tr \phi^2 (0) \rangle, \label{eq:Cmn0} \\
C_2(q) =\left(\frac{N}{g}\right)^2 a^3 \sum_{x \in \Lambda} e^{-i q \cdot x} \langle \Tr \phi^2 (x) \Tr \phi^2 (0)\rangle. \label{eq:C2}
\end{align}
The superscript $0$ is used to distinguish the na\"ively discretized EMT from the renormalized one.

   On the lattice, the correlator $C_\mn(q)$ has a contact term which arises when the operators coincide in position space; in momentum space, this manifests as a constant (momentum-independent) contribution $C_\mn(0)$ which needs to be subtracted before the proper continuum limit can be obtained,
   \begin{align}
   \hat{C}_\mn (q) = C_\mn (q) - C_\mn (0).
   \end{align}
   By dimensional counting, $C_\mn (0)$ has a leading $a^{-1}$ divergent contribution. We therefore define 
   \begin{align} \label{eq:kappa-def}
   C_\mn (0) = \frac{\kappa}{a} \delta_\mn.
\end{align}    
   Lattice perturbation theory at next-to-leading order (NLO) gives the following results for the various expressions from above (details can be found in~\cref{sec:app_c2cmn_0t}):
   \begin{align}
   \hat{C}^{\text{1 loop}}_\mn(q) &= \frac{N^2q}{64}\left(1-\frac{1}{N^2}\right)\pi_\mn +\ord{a},\\
   \hat{C}^{\text{2 loop}}_\mn(q) &=-\frac{N^2q }{256}g_{\text{eff}}\left(1-\frac{1}{N^2}\right)\left(2-\frac{3}{N^2}\right) \pi_\mn +\ord{a},\\
   \kappa &= -\frac{N^2}{2}\left(1-\frac{1}{N^2}\right)\left(\frac{6Z_0-1}{12}\right), \label{eq:kappa-ptval}
   \end{align} 
   where $g_{\text{eff}} = \frac{g}{\vert q \vert}$ is the \textit{effective coupling}, and $\pi_\mn = \delta_\mn - \frac{q_\mu q_\nu}{q^2}$ the \textit{transverse projector}. It can be seen that $\hat{C}_\mn(q)$ has a leading $N^2q$ behavior; an overall $q$ is expected from $\hat{C}_\mn(q)$ being a dimension one correlator, where at LO (i.e. one loop) there is no coupling constant dependence, and at NLO (i.e. two loops) we encounter the first order expansion in the effective coupling $g_{\text{eff}}$. In both terms, the planar diagram contributes to the leading $N^2$ factor, whereas nonplanar diagrams can be seen as $\frac{1}{N^2}$ corrections to the leading planar diagram. The fact that the finite piece of $\hat{C}_\mn(q)$ is proportional to the transverse projector is a consequence of the WI.

  \subsection{Wilson flow}

  From above, we see that the correlator $C^0_\mn(q)$ contains divergent contributions in terms of $\frac{g}{a}c_3$ from the operator mixing , as well as $\frac{\kappa}{a}$ due to the contact term. In order to nonperturbatively renormalize the EMT operator, we need to isolate the contact term from the operator mixing, and we will utilize the method of the Wilson flow~\citep{Luscher:2010iy} to achieve this. For our scalar field $\phi(x)$, define a flowed field $\rho(t,x)$ governed by the flow equations,
\begin{align}
    \partial_t \rho(t,x) = \partial^2 \rho(t,x), \quad \rho(t,x) |_{t=0} = \phi(x),
\end{align}
where $\partial^2 = \sum_\mu \partial_\mu^2$ is the Laplacian, and $t$ is the \textit{flow time}, a new parameter introduced into the theory. Solving by means of Fourier transformation, one finds 
\begin{align}
    \Tilde{\rho}(t, k) = e^{-k^2 t} \Tilde{\phi}(k),
\end{align}
where $\Tilde{\rho}(t,k)$ is the Fourier transform of $\rho(t,x)$; the flow effectively smears the field with radius $\sqrt{4t}$.

The Wilson flow suppresses high-momentum modes exponentially, and thereby regulates the divergent contact term present in the EMT correlator $C^0_\mn (q)$. We are therefore able to isolate the divergent mixing $c_3$ from the divergent contact term. There have been extensive discussions of various implementations of the Wilson flow for renormalizing the EMT, which can be found in~\citep{Suzuki:2013gza,DelDebbio:2013zaa,Capponi:2015ucc,Capponi:2015ahp,Capponi:2016yjz, Giusti:2015daa}.

 In our case, we are interested in determining the flowed correlator 
 \begin{align}
 C_\mn(t, q) =  \frac{N}{g} a^3 \sum_{x \in \Lambda} e^{-i q \cdot x} \langle \Tmn^R (x) \Tr \rho^2 (t, 0) \rangle,
 \end{align}
at finite flow time. Here we replaced the operator $\Tr \phi^2(x=0)$ with the operator $\Tr \rho^2(t, x=0)$ at finite flow time $t$, and kept the renormalized EMT operator $\Tmn^R (x)$ at flow time $t=0$. By definition, $C_\mn(0,q) = C_\mn(q)$. Since the operator mixing $c_3$ is local to the EMT operator $\Tmn(x)$, it is not affected by replacing the probe $\Tr \phi^2(x=0)$ with the one at finite flow time $\Tr \rho^2(t, x=0)$. On the other hand, the divergent contact term $C_\mn(t,q=0)$ is suppressed. More explicitly we similarly define
\begin{align} \label{eq:hat-Cmn}
	\hat{C}_\mn (t, q) &= C_\mn (t, q) - C_\mn (t, 0),\\
	C_\mn(t, 0) &= \delta_\mn K(t).
\end{align}
As recorded in~\cref{eq:kappa-def,eq:kappa-ptval}, at vanishing flow time, $K(t=0) = \frac{\kappa}{a}$. However, as calculated in~\cref{eq:flow-kappa}, at small finite flow time,
\begin{align} \label{eq:flowtime-omega}
K(t) =\frac{\omega}{\sqrt{t}} + \ord{\sqrt{t}},
\end{align}
where at leading order in perturbation theory,
\begin{align}
\omega = -\frac{N^2}{2} \left(1-\frac{1}{N^2}\right)\left(\frac{\sqrt{2}}{24 \pi^{3/2}}\right).
\end{align}

We utilize this small $t$ expansion to remove the contact term contribution in our correlation function in order to obtain the value of $c_3$. The strategy will be explained in further detail in~\cref{sec:renormalisation}.

In analogy to~\cref{eq:Cmn-c3C2,eq:Cmn0,eq:C2} we have the relations 
\begin{align} \label{eq:Cmn-tq}
	C_\mn(t, q)  &= C_\mn^0(t, q) - \frac{g}{a}c_3  \delta_\mn  C_2(t, q),
\end{align} where
\begin{align}
C_\mn^0(t, q) = \frac{N}{g}a^3 \sum_{x \in \Lambda} e^{-i q \cdot x} \langle \Tmn^0 (x) \Tr \rho^2 (t, 0) \rangle, \\
C_2(t, q) =\left(\frac{N}{g}\right)^2 a^3 \sum_{x \in \Lambda} e^{-i q \cdot x} \langle \Tr \phi^2 (x) \Tr \rho^2 (t, 0)\rangle.
\end{align}

Having defined the above correlation functions, we can now nonperturbatively renormalize the EMT on the lattice. The renormalization scheme is defined by first imposing the Ward identity
\begin{align} \label{eq:lattice-WI-renorm-condition}
\bar{q}_\mu \hat{C}_\mn (t, q) = 0
\end{align}
on all lattice ensembles. Here $\bar{q}=\frac{1}{a} \sin \left(aq\right)$ is the lattice momentum. This condition is imposed on specific values of momentum $aq^*$. This gives a value of $c_3$ for each choice of momentum, mass, volume and 't Hooft coupling. We then extrapolate the value $c_3$ towards the massless and infinite volume limit to obtain $\bar{c_3}$. This defines a massless renormalization scheme, which is independent of the volume. We will also investigate the dependence of $c_3$ on the value of the 't Hooft coupling $ag$. The implementation of the scheme and the numerical fits results will be explained in~\cref{sec:renormalisation}.

  \section{Lattice simulations}
  \label{sec:simulations}
    \subsection{Simulation setup}

The theory is simulated using the hybrid Monte Carlo algorithm~\citep{Duane:1987de}, which was implemented using the Grid library~\citep{Boyle:2015tjk, Boyle:2016lbp}. For this paper, we will focus on the $N=2$ theory. The simulated volumes $N_L^3$, 't Hooft coupling in lattice unit $ag$ (or equivalently the dimensionless lattice spacing), and bare masses $(am)^2$ are listed in~\cref{tabel:ensemble}. For each of the three 't Hooft couplings, two bare masses in the vicinity of the critical mass have been simulated (see~\cref{table:critical-mass}).
    
\begin{table}[ht]
    \begin{minipage}{.5\linewidth}
      \centering
        \begin{tabular}{c|c}
        \hline \hline
        	 \quad $ag$  \quad &  \quad $(am)^2$ \quad \\ \hline
             0.1 & -0.0305, -0.031 \\
             0.2 & -0.061, -0.062 \\ 
             0.3 & -0.092, -0.091 \\ 
             \hline \hline
        \end{tabular}
    \end{minipage}%
    \begin{minipage}{.5\linewidth}
      \centering
        \begin{tabular}{c|c|c}
        \hline \hline
             \quad $N_L^3$  \qquad & \quad Trajectories \quad & \quad Sample frequency \\ \hline
            $64^3$ & 1,500,000 & 50\\ 
            $128^3$& 500,000 & 50\\
            $256^3$ & 200,000 & 100\\
            \hline \hline
        \end{tabular}
    \end{minipage} 
    \caption{For each 't Hooft coupling $ag$, two bare masses are simulated in three volumes}
    \label{tabel:ensemble}
\end{table}

\begin{figure}[h]
  \includegraphics[scale=1.0]{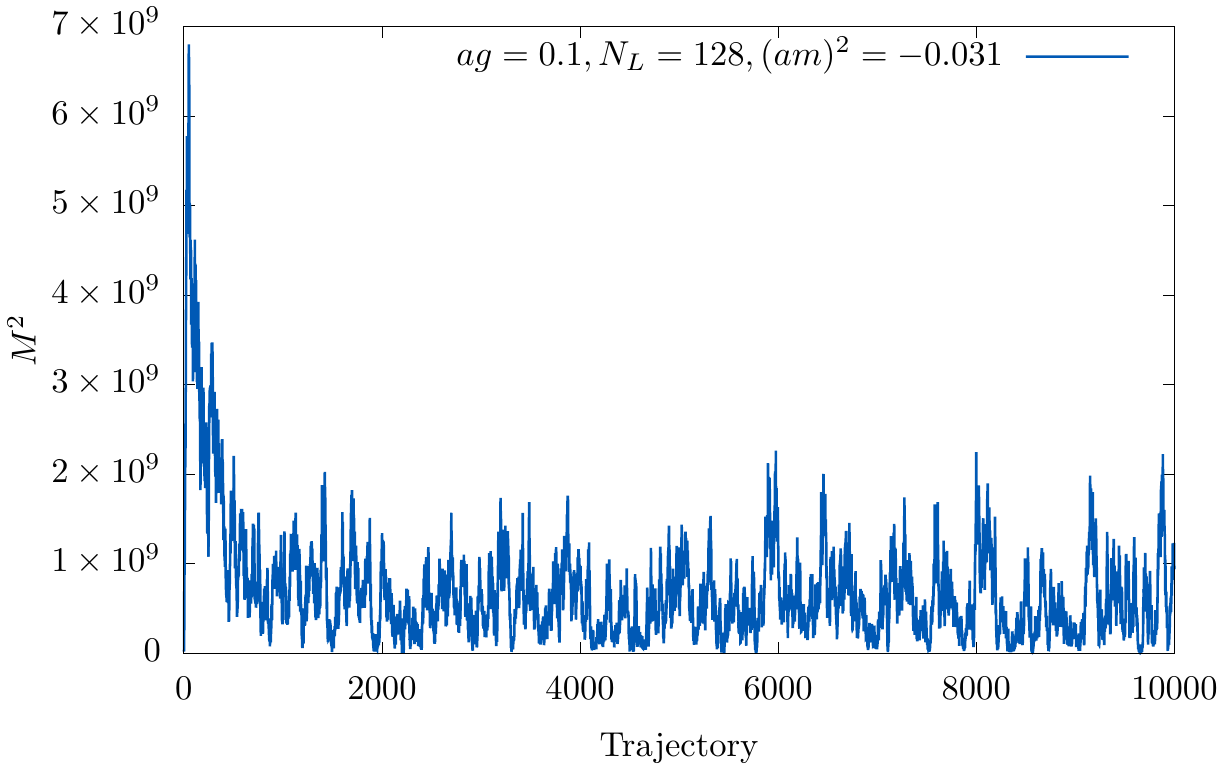}
  \caption{Example of observable $M^2= \Tr \left(a^3 \suml{x} \phi(x) \right)^2$ for ensemble with $ag=0.1, N_L=128, (am)^2=-0.031$}
  \label{fig:traj-autocorrelation}
\end{figure}

Correlation function computations are performed using the Hadrons library~\citep{antonin_portelli_2020_4293902} and the data analysis is based on the LatAnalyze library~\citep{antonin_portelli_2020_4293639}. The data and analysis code are available at~\citep{joseph_k_l_lee_2020_4293949,joseph_k_l_lee_2020_4293702,del_debbio_luigi_2020_4290392}. Data analysis is performed using bootstrap resampling~\citep{Efron:1979bxm}, and only every 50th or 100th trajectory is sampled in order to reduce autocorrelation. The first 5000 trajectories are discarded to ensure the ensembles are thermalized. A representative example of the value of the observable $M^2= \Tr \left( a^3 \suml{x} \phi(x) \right)^2$ across one HMC simulation ($ag=0.1, N_L=128, (am)^2=-0.031$) is shown in~\cref{fig:traj-autocorrelation}.

  \subsection{Critical mass determination}
To extrapolate to the massless point, the renormalized masses of the ensembles have to be determined, which requires the critical masses for each lattice spacing as input. These have been determined in~\citep{Cossu:2020yeg,cossu_guido_2020_4266114,kitching_morley_ben_2020_4290508} at two loops in lattice perturbation theory, as well as nonperturbatively by analyzing the finite-size scaling of the Binder cumulant. The relevant masses are summarized in~\cref{table:critical-mass}. 
\begin{table}[h]
\centering
\begin{tabular}{c|c|c|c}
\hline \hline
 \quad $ag$  \quad &  \quad One loop  \quad &  \quad Two loop  \quad &  \quad Nonperturbative  \quad \\  \hline
 0.1 &  -0.03159 &  -0.03125&   -0.0313408(38) \\  \hline
 0.2 & -0.06318 &  -0.06194 & -0.0622974(98) \\  \hline
 0.3 & -0.09477 & -0.09208  &  -0.092935(16)    \\
 \hline \hline
  \end{tabular}
  \caption{The critical masses $(am_c)^2$ in the infinite volume limit are calculated at NLO in lattice perturbation theory, as well as nonperturbatively in~\citep{Cossu:2020yeg}, which are listed for each 't Hooft coupling $ag$. These are used in the later global fit to obtain $c_3$ in the massless limit.}
  \label{table:critical-mass}
\end{table}
  
  \section{Renormalization of the EMT}
  \label{sec:renormalisation}
The renormalisation condition~\cref{eq:lattice-WI-renorm-condition} implies that $\hat{C}_\mn (t,q)$ is purely transverse, \ie, 
\begin{align} \label{eq:Cmntq-transverse}
\hat{C}_\mn (t,q) = F(t, q)\bar{\pi}_\mn
\end{align}
where $\bar{\pi}_\mn = \delta_\mn - \frac{\bar{q}_\mu \bar{q}_\nu}{\bar{q}^2}$ is the transverse projector with lattice momentum $\bar{q}$. In other words, $\hat{C}_\mn$ vanishes in the direction with purely longitudinal momentum. For example, picking the momentum to be purely in the direction $q_l=(q_0, q_1, q_2)=(0, 0, q_2)$, 
\begin{align}
\hat{C}_{22} (t, q_l) = 0.
\end{align}    

Substituting the definition of $\hat{C}_\mn(t,q_l)$ from~\cref{eq:hat-Cmn,eq:Cmn-tq}, we obtain
\begin{align}
\hat{C}_{22} (t, q_l) &= C_{22}(t,q_l) - C_{22}(t,0) = C^0_{22}(t, q_l) - \frac{g}{a}c_3 C_2(t,q_l) - K(t) = 0\\
\to c_3 &= \frac{a}{g} \frac{C^0_{22}(t,q_l)}{C_2(t,q_l)} - f_g(g\sqrt{t}, q_l), \label{eq:c3-flow}
\end{align}
where
\begin{align} \label{eq:fg-ratio}
f_g(g\sqrt{t}, q_l) = \frac{a}{g}  \frac{K(t)} {C_2(t,q_l)}.
\end{align}
Using the one-loop perturbative expressions for $K(t)$ and $C_2(t,q)$ from~\cref{eq:append-flow-c2-expansion,eq:flow-kappa}, this gives
\begin{align} \label{eq:flow-fg}
f_g(g\sqrt{t}, q_l) = \frac{\omega'(q_l)}{g\sqrt{t}} + \ord{q_l\sqrt{t}},
\end{align} 
where $\omega'(q) = \frac{\sqrt{2} (aq)}{3 \pi^{3/2}}$. (Details can be found in~\cref{sec:app_c2cmn_ft}). The strategy to obtain the value of $c_3$ is to first flow the correlators to a range of small finite flow times, at a fixed momentum $aq^*_l$. Then, utilizing~\cref{eq:c3-flow}, we fit the ratio on the left-hand side of
\begin{align} 
\frac{a}{g} \frac{C^0_{22}(t,q^*_l)}{C_2(t,q^*_l)} = c_3 + f_g(g\sqrt{t},q^*_l),
\end{align}
as a function of the \textit{physical flow time} $g\sqrt{t}$. We have tested a range of fit functions for $f_g$, and have found that the fit ansatz
\begin{align}\label{eq:c3-fit-equation}
    \frac{a}{g} \frac{C^0_{22}(t,q^*_l)}{C_2(t,q^*_l)} = c_3 + \frac{\Omega}{g\sqrt{t}}
\end{align}
provides a very good fit to the data. Here we keep the first term linear in the \textit{inverse physical flow time} $\frac{1}{g\sqrt{t}}$ from~\cref{eq:flow-fg}, and leave $\Omega$ and $c_3$ as fit parameters. From the fit we can extrapolate $c_3$ from the $y$ intercept.

        \subsection{Numerical results}

    Picking the fit ranges for the physical flow time $g\sqrt{t}$ requires special attention. They must first be sufficiently small to justify the small flow time expansion of~\cref{eq:flow-fg}. This also ensures the smearing radius is sufficiently smaller than the length of the lattice ($gL=gaN_L$) such that there will be small finite volume contributions from the boundaries. The physical flow time must also be larger than the lattice spacing ($ag$) such that actual smearing occurs across lattice points. We therefore impose the range to be between $ag < g \sqrt{t} < 1$. We performed the analysis for four values of momenta $a\vert q^*_l\vert = 0.049$, $0.098$, $0.147$, $0.196$.

\begin{figure}[h]
\centering
\subfigure[\ $ag=0.1, N_L=64$]{
\includegraphics[scale=0.84]{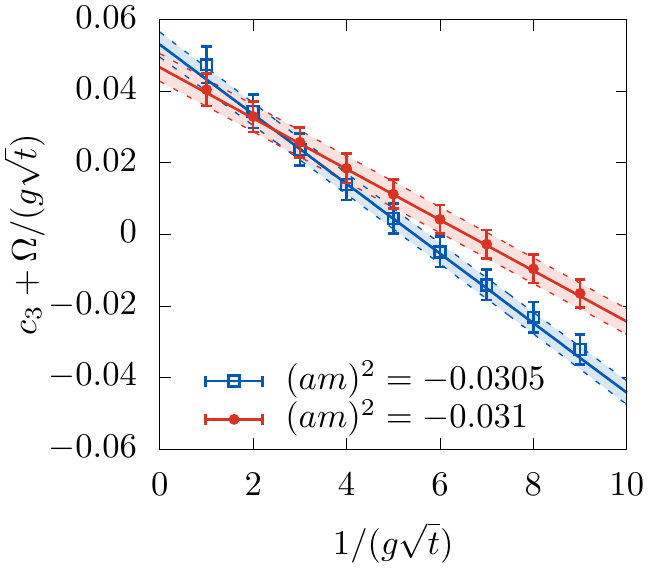}}
\subfigure[\ $ag=0.1, N_L=128$]{
\includegraphics[scale=0.84]{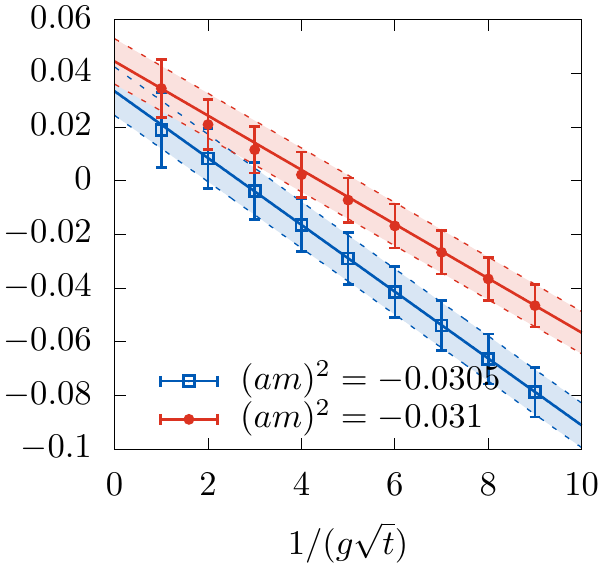}}
\subfigure[\ $ag=0.1, N_L=256$]{
\includegraphics[scale=0.84]{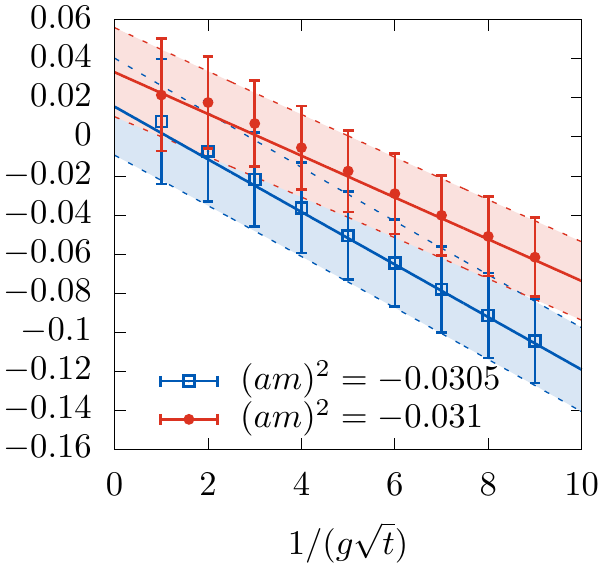}}
\centering
\subfigure[\ $ag=0.2, N_L=64$]{
\includegraphics[scale=0.84]{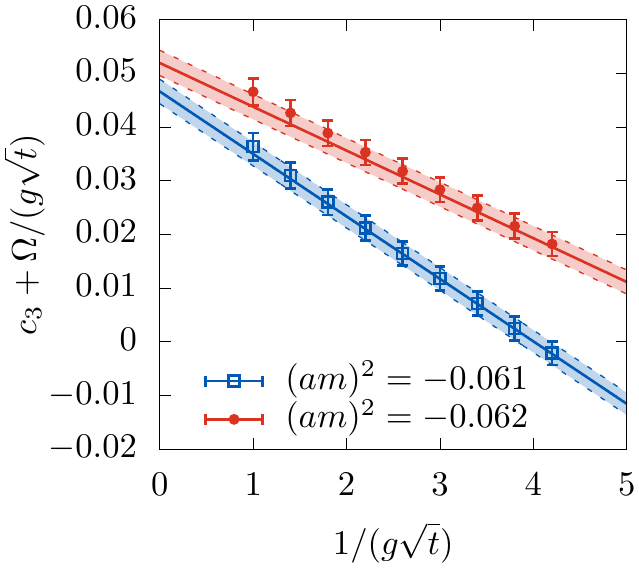}}
\subfigure[\ $ag=0.2, N_L=128$]{
\includegraphics[scale=0.84]{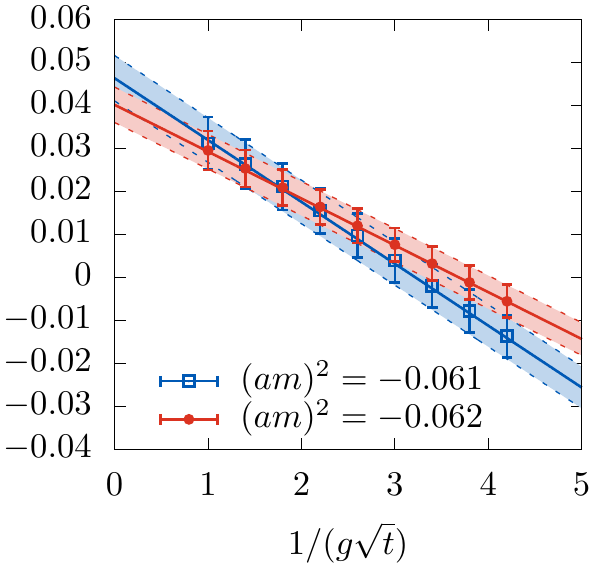}}
\subfigure[\ $ag=0.2, N_L=256$]{
\includegraphics[scale=0.84]{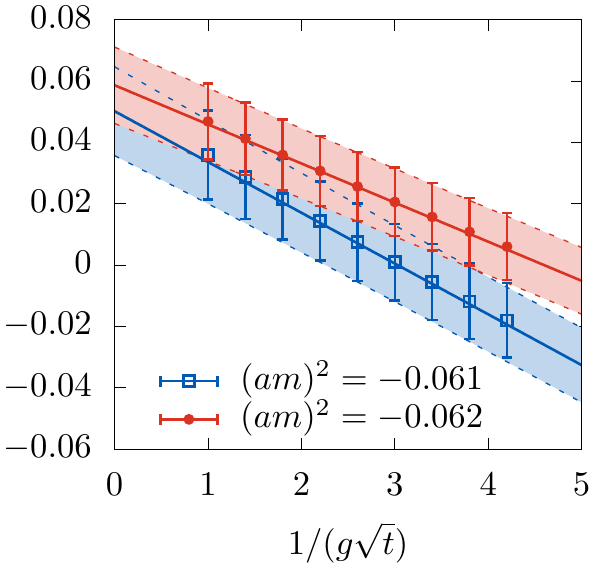}}
\centering
\subfigure[\ $ag=0.3, N_L=64$]{
\includegraphics[scale=0.84]{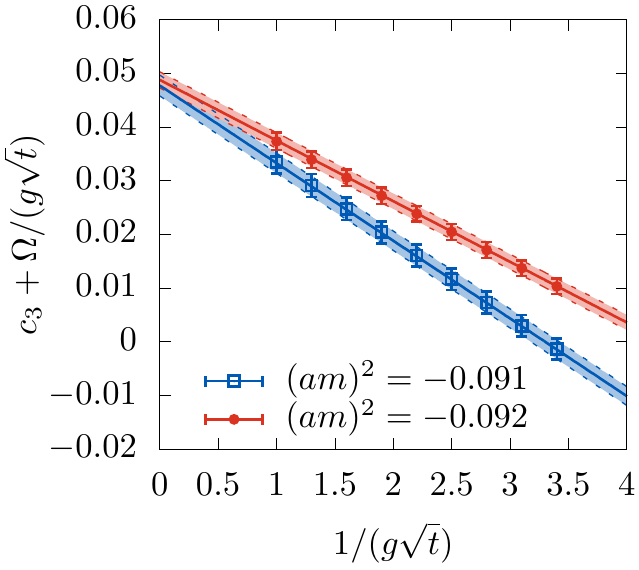}}
\subfigure[\ $ag=0.3, N_L=128$]{
\includegraphics[scale=0.84]{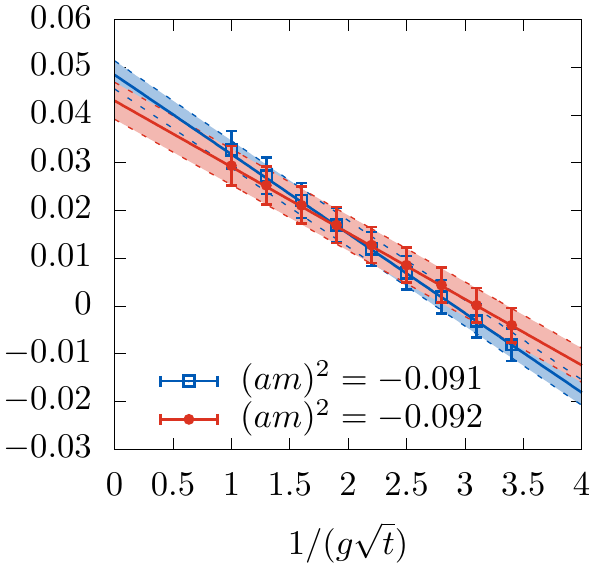}}
\subfigure[\ $ag=0.3, N_L=256$]{
\includegraphics[scale=0.84]{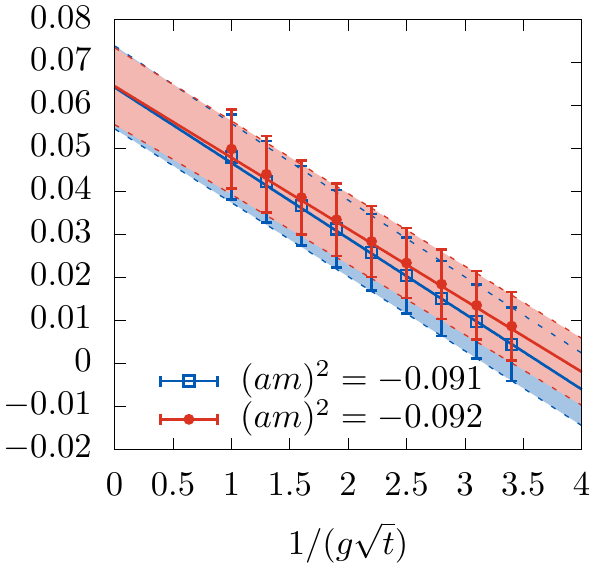}}
\caption{Plots showing $c_3$ against the inverse physical flow time $\frac{1}{g\sqrt{t}}$ using~\cref{eq:c3-fit-equation} for three 't Hooft couplings and three volumes at momentum $a\vert q^*_l \vert = 0.098$. The red and blue data points are for the lighter and heavier mass simulations respectively, and the corresponding error bands in the fit are from statistical uncertainty. The value of $c_3$ is the $y$ intercept on the fit.\label{fig:c3fit-plot}}
\end{figure}

The fits with respect to the inverse flow time for one of the momenta $a\vert q^*_l \vert = 0.098$ are shown in~\cref{fig:c3fit-plot}, and the fit values of $c_3$ for each ensemble are summarized in~\cref{table:c3-ensemble}.

    \begin{table}[h]
    \centering
\begin{tabular}{c|c|c|c|c|c|c|c}
\hline \hline
 \quad $ag$  \quad & \quad  $N_L$  \quad & \quad  $(am)^2$  \quad & \quad  $a\vert q^*_l \vert$  \quad & \quad  dof  \quad & \quad  $\chi^2/dof$  \quad & \quad  $p$ value  \quad & \quad  $c_3$         \\ \hline
0.1  & 64    & -0.0305  & 0.098                 & 4   & 1.60         & 0.17                         & 0.0531(35) \\
0.1  & 64    & -0.031   & 0.098                 & 3   & 0.15         & 0.93                         & 0.0467(39) \\
0.1  & 128   & -0.0305  & 0.098                 & 5   & 0.06         & 1.00                         & 0.0334(90) \\
0.1  & 128   & -0.031   & 0.098                 & 5   & 1.00         & 0.42                         & 0.0445(85) \\
0.1  & 256   & -0.0305  & 0.098                 & 3   & 0.18         & 0.91                         & 0.015(25)  \\
0.1  & 256   & -0.031   & 0.098                 & 3   & 1.26         & 0.28                         & 0.033(23)  \\ \hline
0.2  & 64    & -0.061   & 0.098                 & 5   & 1.14         & 0.34                         & 0.0466(23) \\
0.2  & 64    & -0.062   & 0.098                 & 5   & 1.64         & 0.15                         & 0.0519(24) \\
0.2  & 128   & -0.061   & 0.098                 & 5   & 0.70         & 0.62                         & 0.0464(53) \\
0.2  & 128   & -0.062   & 0.098                 & 5   & 0.67         & 0.65                         & 0.0402(41) \\
0.2  & 256   & -0.061   & 0.098                 & 2   & 0.27         & 0.77                         & 0.050(14)  \\
0.2  & 256   & -0.062   & 0.098                 & 2   & 0.04         & 0.96                         & 0.059(12)  \\ \hline
0.3  & 64    & -0.091   & 0.098                 & 5   & 0.56         & 0.73                         & 0.0478(19) \\
0.3  & 64    & -0.092   & 0.098                 & 5   & 0.85         & 0.51                         & 0.0488(15) \\
0.3  & 128   & -0.091   & 0.098                 & 6   & 0.52         & 0.79                         & 0.0484(29) \\
0.3  & 128   & -0.092   & 0.098                 & 5   & 0.85         & 0.52                         & 0.0430(39) \\
0.3  & 256   & -0.091   & 0.098                 & 3   & 0.14         & 0.94                         & 0.0643(97) \\
0.3  & 256   & -0.092   & 0.098                 & 3   & 0.52         & 0.67                         & 0.0645(89) \\
\hline \hline
  \end{tabular}
  \caption{For each simulation, we perform the fit for the value of $c_3$ using~\cref{eq:c3-fit-equation}. This table shows the fits for momentum $a\vert q^*_l \vert = 0.098$, and the flow time fit range is bounded by $ag < g \sqrt{t} < 1$.}
  \label{table:c3-ensemble}
\end{table}

In order to include the mass, volume and lattice-spacing dependence of the value of $c_3$, we perform global fits using
\begin{align} \label{eq:global-fit-model}
    c_3(\overbar{m^2_R},gL, ag) = \overbar{c_3} + p_0 \bar{m^2_R} + \frac{p_1}{gL} + p_2 (ag),
\end{align}
where $\overbar{m^2_R}=(m^2-m_c^2)/g^2$ is the dimensionless renormalized mass (The values of $m_c^2$ are summarized in~\cref{table:critical-mass}), $gL$ is the dimensionless length of the lattice, and $ag$ the dimensionless lattice spacing. As we have chosen our simulation to have large volume, small lattice spacing, and close to the critical mass, we believe that the linear corrections are appropriate. In particular, since the divergent mixing is a UV effect, we expect there to be small volume dependence coming from the IR. 

\begin{figure}[hp]
\centering
  \includegraphics[scale=1.0]{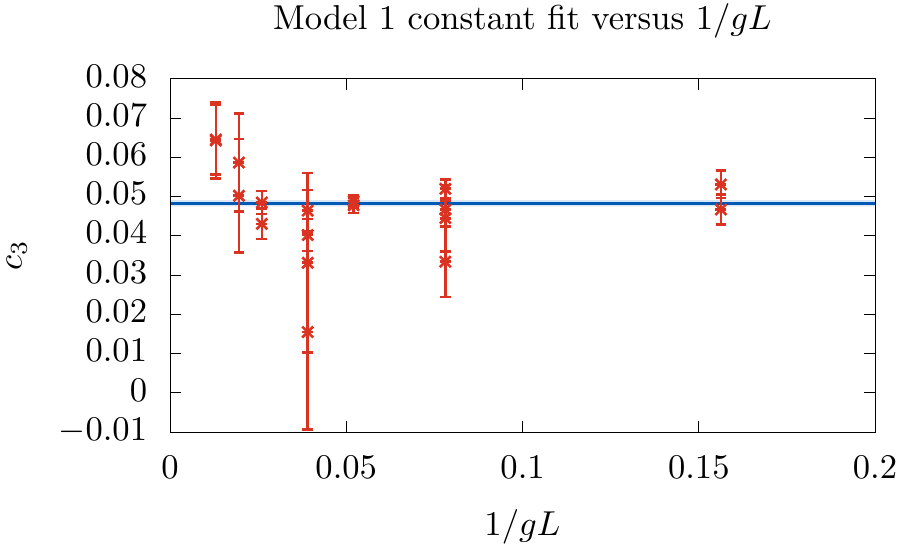}
  \caption{$c_3$ global fit using model 1 for $a \vert q^*_l \vert = 0.098$. The value of $c_3$ is plotted against $\frac{1}{gL}$. \label{fig:global-fit-model-1}}
\end{figure}

\begin{figure}[hp]
\centering
\subfigure[\ Model 2 (against $\bar{m_R^2}$)]{
\includegraphics[scale=1.0]{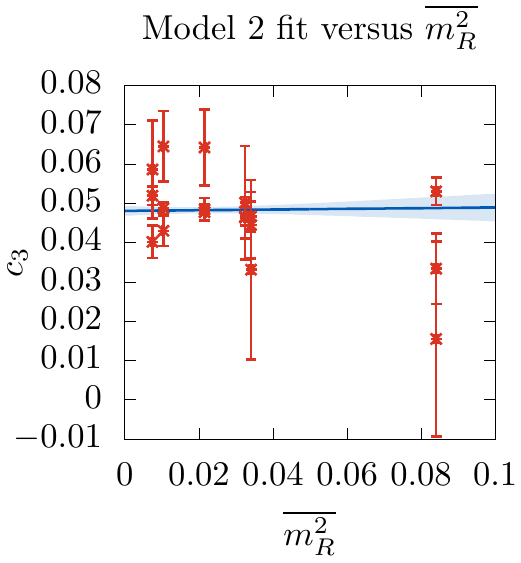}}
\subfigure[\ Model 3 (against $gL$)]{
\includegraphics[scale=1.0]{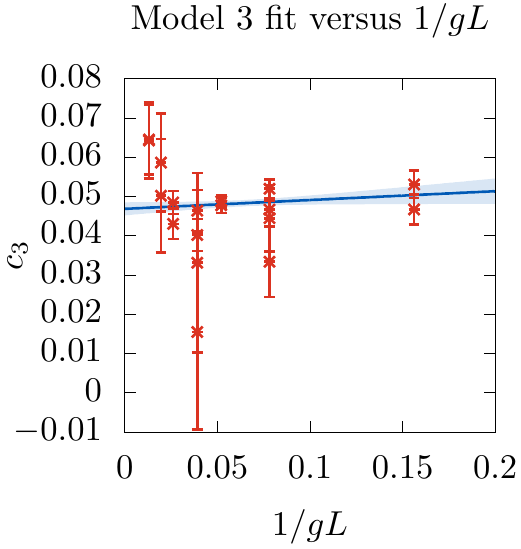}}
\subfigure[\ Model 4 (against $ag$)]{
\includegraphics[scale=1.0]{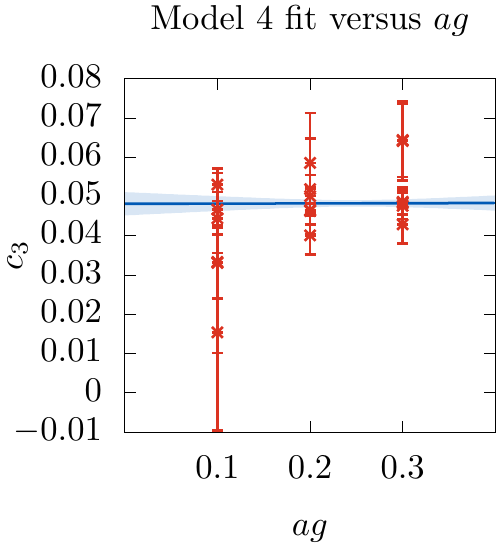}}
\caption{$c_3$ global fits using model 2, 3, 4 for $a \vert q^*_l \vert = 0.098$.. Each plot is plotted against the respective free fitting parameter for each model, \ie $\bar{m_R^2}$, $gL$, and $ag$; the value for $\overbar{c_3}$ is the $y$-intercept of the fit line.\label{fig:global-fits}}
\end{figure}

For the global fits, the three parameters $p_0$, $p_1$, $p_2$ are switched on individually, resulting in $2 \times 2 \times 2 = 8$ fit models for each of the four momenta, which gives a total of 32 fit results for the value of $\overbar{c_3}$. The fit values for $\overbar{c_3}$ using different models are summarized in~\cref{table:global-fit}.  \cref{fig:global-fit-model-1,fig:global-fits} show examples of the global fits for models 1-4 for momentum $a \vert q^*_l \vert = 0.098$.


\begin{table}[hp]
\centering
\begin{tabular}{c|c|c|c|c|c|c|c|c}
\hline \hline
 \quad $a \vert q^*_l \vert$  \quad & \quad Model  \quad &  \quad Fit parameters  \quad                               &  \quad $\overbar{c_3}$   \quad  &  \quad $p_0$     \quad    & \quad  $p_1$      \quad & \quad  $p_2$        \quad  & $\chi^2/$dof  \quad & \quad  $p$-value  \quad \\ \hline
0.049 & 1 & $\overbar{c_3}$               & 0.0489(15)  &            &            &            & 0.49 & 0.18  \\
0.049 & 2 & $\overbar{c_3}, p_0$          & 0.0514(24)  & -0.129(93) &            &            & 0.35 & 0.06  \\
0.049 & 3 & $\overbar{c_3}, p_1$          & 0.0499(36)  &            & -0.029(96) &            & 0.53 & 0.26  \\
0.049 & 4 & $\overbar{c_3}, p_2$          & 0.0456(58)  &            &            & 0.014(24)  & 0.5  & 0.22  \\
0.049 & 5 & $\overbar{c_3},p_0, p_1$      & 0.0498(36)  & -0.17(11)  & 0.07(12)   &            & 0.35 & 0.08  \\
0.049 & 6 & $\overbar{c_3},p_1, p_2$      & 0.040(16)   &            & 0.07(18)   & 0.031(44)  & 0.55 & 0.32  \\
0.049 & 7 & $\overbar{c_3},p_0, p_2$      & 0.0522(78)  & -0.13(11)  &            & -0.003(27) & 0.39 & 0.11  \\
0.049 & 8 & $\overbar{c_3},p_0, p_1, p_2$ & 0.041(16)   & -0.17(11)  & 0.16(19)   & 0.026(44)  & 0.35 & 0.11  \\ \hline 
0.098 & 1 & $\overbar{c_3}$               & 0.04828(81) &            &            &            & 1.4  & 0.25  \\
0.098 & 2 & $\overbar{c_3}, p_0$          & 0.0481(12)  & 0.009(44)  &            &            & 1.49 & 0.19  \\
0.098 & 3 & $\overbar{c_3}, p_1$          & 0.0469(17)  &            & 0.022(23)  &            & 1.43 & 0.23  \\
0.098 & 4 & $\overbar{c_3}, p_2$          & 0.0481(30)  &            &            & 0.000(12)  & 1.49 & 0.19  \\
0.098 & 5 & $\overbar{c_3},p_0, p_1$      & 0.0468(17)  & -0.038(60) & 0.036(32)  &            & 1.5  & 0.19  \\
0.098 & 6 & $\overbar{c_3},p_1, p_2$      & 0.0363(74)  &            & 0.072(41)  & 0.030(20)  & 1.38 & 0.3   \\
0.098 & 7 & $\overbar{c_3},p_0, p_2$      & 0.0471(48)  & 0.017(58)  &            & 0.003(15)  & 1.58 & 0.14  \\
0.098 & 8 & $\overbar{c_3},p_0, p_1, p_2$ & 0.0368(76)  & -0.018(62) & 0.076(43)  & 0.029(21)  & 1.47 & 0.22  \\ \hline 
0.147 & 1 & $\overbar{c_3}$               & 0.0418(23)  &            &            &            & 0.98 & 0.93  \\
0.147 & 2 & $\overbar{c_3}, p_0$          & 0.0445(41)  & -0.14(18)  &            &            & 1.01 & 0.86  \\
0.147 & 3 & $\overbar{c_3}, p_1$          & 0.0496(60)  &            & -0.25(18)  &            & 0.88 & 0.9   \\
0.147 & 4 & $\overbar{c_3}, p_2$          & 0.029(10)   &            &            & 0.051(39)  & 0.91 & 0.95  \\
0.147 & 5 & $\overbar{c_3},p_0, p_1$      & 0.0497(61)  & -0.04(20)  & -0.24(20)  &            & 0.97 & 0.92  \\
0.147 & 6 & $\overbar{c_3},p_1, p_2$      & 0.042(25)   &            & -0.18(30)  & 0.020(65)  & 0.97 & 0.93  \\ 
0.147 & 7 & $\overbar{c_3},p_0, p_2$      & 0.031(13)   & -0.04(20)  &            & 0.046(44)  & 1    & 0.88  \\
0.147 & 8 & $\overbar{c_3},p_0, p_1, p_2$ & 0.043(25)   & -0.03(20)  & -0.17(31)  & 0.018(66)  & 1.09 & 0.74  \\ \hline 
0.196 & 1 & $\overbar{c_3}$               & 0.0414(24)  &            &            &            & 0.33 & 0.08  \\
0.196 & 2 & $\overbar{c_3}, p_0$          & 0.0476(50)  & -0.36(26)  &            &            & 0.09 & 0.002 \\
0.196 & 3 & $\overbar{c_3}, p_1$          & 0.0452(57)  &            & -0.067(91) &            & 0.29 & 0.09  \\
0.196 & 4 & $\overbar{c_3}, p_2$          & 0.034(11)   &            &            & 0.028(42)  & 0.31 & 0.1   \\
0.196 & 5 & $\overbar{c_3},p_0, p_1$      & 0.0485(63)  & -0.34(28)  & -0.024(97) &            & 0.1  & 0.01  \\
0.196 & 6 & $\overbar{c_3},p_1, p_2$      & 0.042(24)   &            & -0.05(14)  & 0.009(65)  & 0.34 & 0.17  \\
0.196 & 7 & $\overbar{c_3},p_0, p_2$      & 0.046(15)   & -0.35(28)  &            & 0.007(45)  & 0.11 & 0.01  \\
0.196 & 8 & $\overbar{c_3},p_0, p_1, p_2$ & 0.050(24)   & -0.34(28)  & -0.03(14)  & -0.004(65) & 0.12 & 0.02 \\
\hline \hline
\end{tabular}
\caption{Each global fit model is defined by including a combination of the parameters $(p_0, p_1, p_2)$ from~\cref{eq:global-fit-model} along with the value of $\overbar{c_3}$. For reference, the 1-loop perturbative value,~\cref{eq:c3-1loop}, gives $c_3^{\text{1 loop}} \approx 0.05379$.}
\label{table:global-fit}
\end{table}

\begin{figure}[h]
  \centering
  \includegraphics[scale=1.0]{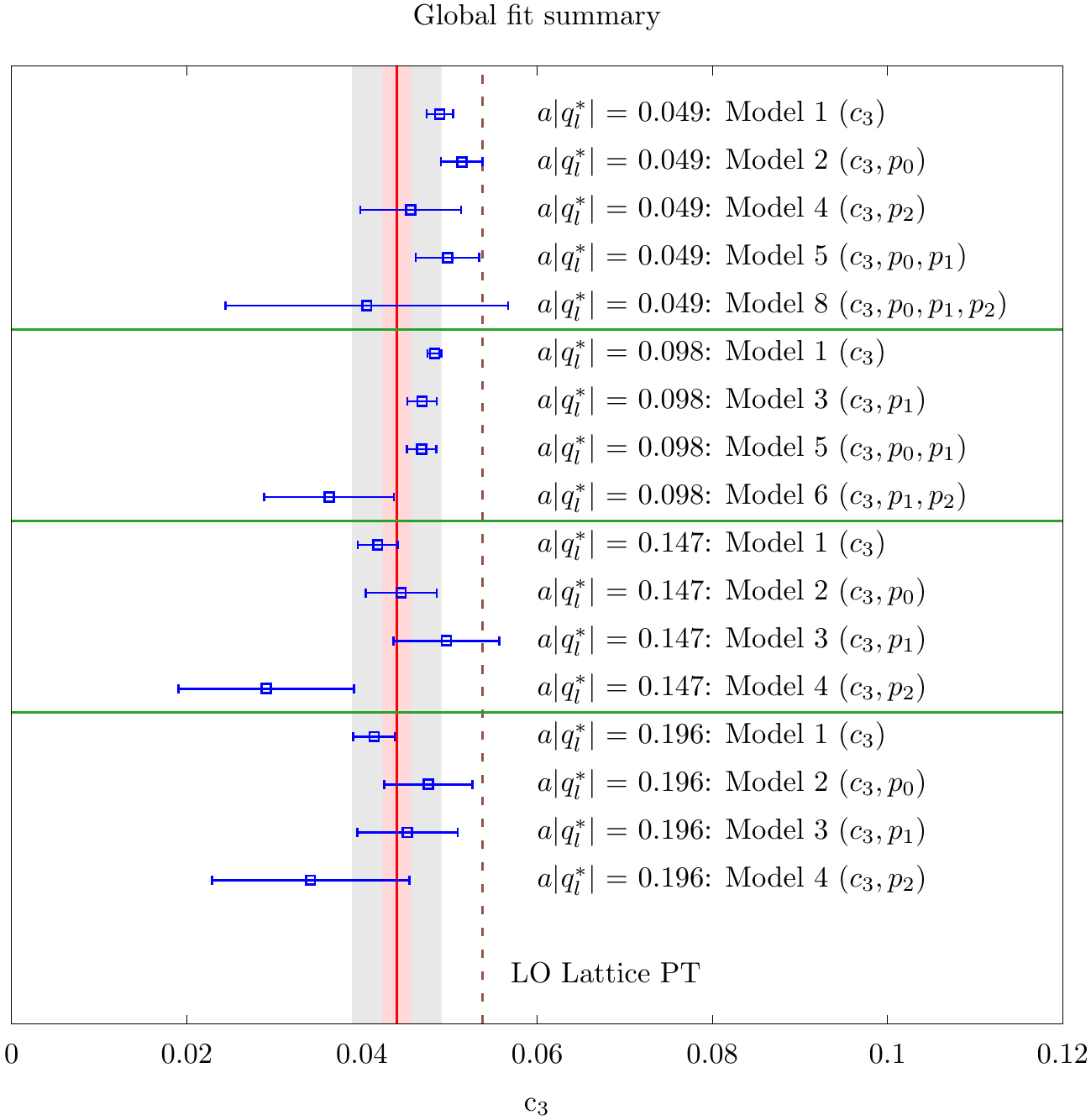}
  \caption{The values of $\overbar{c_3}$ from~\cref{table:global-fit} for models with no fit parameters which are $0.5 \sigma$ compatible with 0. The red line shows the final central result, the red and grey bands represent the statistical and systematic errors respectively. The brown dashed line shows the 1-loop perturbation theory value from~\cref{eq:c3-1loop}. The fact that the nonperturbative result is close to the 1-loop perturbative value is expected due to the super-renormalizability of the theory.}
  \label{fig:global-fit-summary}
\end{figure}

\begin{figure}[h]
  \centering
  \includegraphics[scale=1.0]{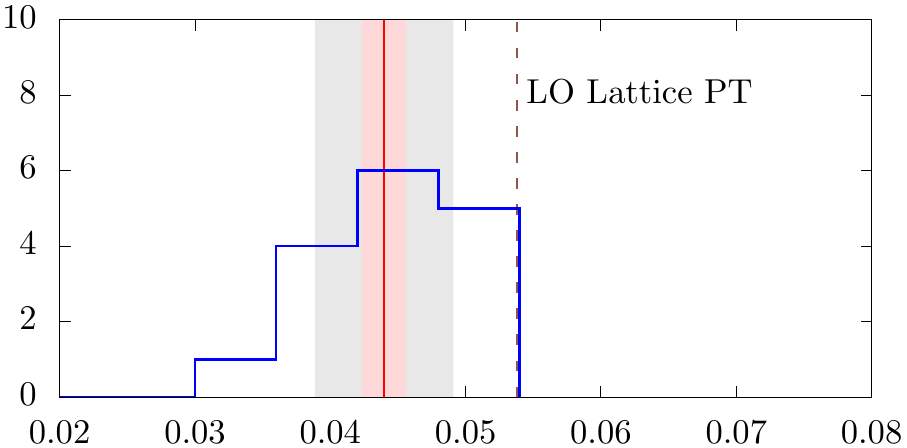}
  \caption{Histogram of the distribution in~\cref{fig:global-fit-summary}. The red line shows the final central result, the red and gray bands represent the statistical and systematic errors respectively. The brown dashed line shows the one-loop perturbation theory value from~\cref{eq:c3-1loop}. The fact that the nonperturbative result is close to the one-loop perturbative value is expected due to the super-renormalizability of the theory.}
  \label{fig:global-fit-histogram}
\end{figure}

In order to estimate the final statistical and systematic errors, we adopt the following procedure inspired by~\citep{Durr:2008zz}. We construct the distribution of values for $\overbar{c_3}$ from global fits which does not include any parameter with a fit value $0.5 \sigma$ compatible with 0. From the 17 results within the distribution, the central value of $\overbar{c_3}$ is defined to be the mean of the distribution, the statistical error to be the statistical error of the mean as measured with the bootstrap samples, and the systematic error to be the symmetrized central 68.3\% confidence interval of the distribution.
A summary of the values of $\overbar{c_3}$ and a histogram of the distribution are shown in~\cref{fig:global-fit-summary,fig:global-fit-histogram} respectively, along with the one-loop value $c_3^{\text{1 loop}}$ from~\cref{eq:c3-1loop}. This procedure yields the final result $\overbar{c_3} = 0.0440(16)_\text{stat}(51)_\text{sys}$.

It is worth noting again that the finiteness of this value in the infinite volume limit is a nonpeturbative feature of the theory. In perturbation theory, all terms of $\ord{g^2}$ are IR divergent and depend on the IR regulator; but as shown in~\citep{Cossu:2020yeg} the theory is in fact nonperturbatively IR finite, where the dimensionful coupling effectively acts as the IR regulator in the infinite volume limit. Comparing the nonperturbative result for $\overbar{c_3}$ with the one-loop perturbative value, the nonperturbative value is approximately $20\%$ smaller than the one-loop result. This is qualitatively expected, as the higher order terms in perturbation theory (with the IR regulator replaced by the coupling) changes sign at every order, and the two-loop result is a correction of the opposite sign to the one-loop value.


  \section{Conclusion and outlook}
  \label{sec:conclusion}
We have presented a procedure to nonperturbatively renormalize the EMT on the lattice for a three-dimensional scalar QFT with a $\varphi^4$ interaction and field $\varphi$ in the adjoint of $SU(N)$. We have also presented numerical results of the EMT operator mixing for the theory with $N=2$. The method utilizes the Wilson flow to define a probe at positive flow time, which can eliminate the divergent contact term present in the EMT correlator. This allows us to determine the mixing coefficient with the lower-dimensional operator $\delta_\mn\frac{N}{g}\Tr \phi^2$. This ensures that the Ward identity can be restored in the continuum limit, up to cutoff effects.

The context of our investigation is to predict the CMB power spectrum for holographic cosmological models, and to test them against observational data. The next step of the investigation is to determine the renormalized EMT two-point function, $C_\mnrs (q) = \langle T_\mn(q) T_\rs(-q) \rangle$, for this class of scalar theories. This two-point function can be used to compute the primordial CMB power spectra in the holographic cosmology framework. On the lattice, this correlator contains a large contact term of order $\ord {a^{-3}}$. This large contact term presents significant statistical noise to the signal of the renormalized two-point function. We are currently exploring using the Wilson flow to eliminate the presence of such a contact term, which will allow us to make a fully nonperturbative prediction for the CMB power spectra with the $SU(N)$ scalar theory as the dual theory.

We are also working towards simulating and performing the renormalization of the EMT for three-dimensional QFTs with adjoint $SU(N)$ scalars coupled to gauge fields. This is the class of theories preferred by the fit of the perturbative predictions to Planck data~\citep{Afshordi:2016dvb}, and has been extensively studied in the literature~\citep{Laine:1995np,Kajantie:1995dw,Laine:1997dy,Kajantie:1997tt,Kajantie:2003ax,Hietanen:2008tv}. In these theories, the lattice EMT contains more counterterms which need to be determined. Much work has been performed in studying the EMT on the lattice for gauge theories~\citep{Caracciolo:1991cp} and gauge theories with fermions~\citep{Caracciolo:1989pt}. The implementation of the Wilson flow for renormalizing the EMT has also been studied for gauge theories~\citep{Suzuki:2013gza,Asakawa:2013laa,DelDebbio:2013zaa, Makino:2014wca,Makino:2014taa,Kitazawa:2014uxa,Capponi:2015ahp,Kitazawa:2016dsl,Kitazawa:2017qab,Hirakida:2018uoy}. We are exploring related methods to perform renormalization of the EMT for theories with scalar fields coupled to gauge fields. This will take us closer to fully testing the viability of holographic cosmological models as a description of the very early Universe.
The supporting data for this article are openly available from~\citep{joseph_k_l_lee_2020_4293949,joseph_k_l_lee_2020_4293702,del_debbio_luigi_2020_4290392}. 
  \begin{acknowledgments}
    The authors would like to warmly thank Pavlos Vranas for his valuable support during the early stages of this project. We thank Masanori Hanada for collaboration at initial stages of this project. Simulations produced for this work were performed using the Grid~\citep{Boyle:2015tjk, Boyle:2016lbp} and Hadrons~\citep{antonin_portelli_2020_4293902} libraries, which are free software under GPLv2. The data analysis was based on the LatAnalyze library~\citep{antonin_portelli_2020_4293639}, which is free software under GPLv3. This work was performed using the Cambridge Service for Data Driven Discovery (CSD3), part of which is operated by the University of Cambridge Research Computing on behalf of the STFC DiRAC HPC Facility~\citep{Diracsite}. The DiRAC component of CSD3 was funded by BEIS capital funding via STFC capital grants ST/P002307/1 and ST/R002452/1 and STFC operations grant ST/R00689X/1. DiRAC is part of the National e-Infrastructure. A.J. and K.S. acknowledge funding from STFC consolidated grants ST/P000711/1 and ST/T000775/1. A.P. is supported in part by UK STFC grant ST/P000630/1. A.P., J.K.L.L., V.N., and H.B.R are funded in part by the European Research Council (ERC) under the European Union’s Horizon 2020 research and innovation programme under grant agreement No 757646 and A.P. additionally grant agreement No 813942. J.K.L.L. is also partly funded by the Croucher foundation through the Croucher Scholarships for Doctoral Study. B.K.M. was supported by the EPSRC Centre for Doctoral Training in Next Generation Computational Modelling Grant No. EP/L015382/1. V.N. is partially funded by the research internship funds of the Universit\'e Paris-Saclay. L.D.D. is supported by an STFC Consolidated Grant, ST/P0000630/1, and a Royal Society Wolfson Research Merit Award, WM140078. 

  \end{acknowledgments}
  \appendix*
  \section{Lattice perturbation theory calculations}
  \label{sec:appendixA}
In this appendix we present the details of the lattice perturbation theory (LPT) calculations in~\cref{sec:generalities}. We will first evaluate two lattice scalar integrals in~\cref{sec:app_eval_ints}, which are necessary to calculate the EMT $c_3$ coefficient mixing in~\cref{sec:app_emt_c3}, the correlators $C_2(q)$, $C_\mn(q)$ at vanishing flow time in~\cref{sec:app_c2cmn_0t}, and the correlators $C_2(t, q)$, $C_\mn(t, q)$ at finite flow time in~\cref{sec:app_c2cmn_ft}.

\subsection{Massless lattice integrals: $V(q)$ and $I_\mn (q)$}\label{sec:app_eval_ints}
		To evaluate the relevant massless lattice integrals, we generalize the method used in~\citep{Burgio:1996ji} to three dimensions. Using a set of recursion relations, any massless, one-loop lattice scalar integrals in three dimensions of the form
    \begin{equation}
      \cB_\epsilon(p;n) = \lim_{\delta\to 0}\int_{-\pi/a}^{\pi/a}\frac{\d^3 k}{(2\pi)^3}
      \;\frac{\hat k_0^{2n_0}\hat k_1^{2n_1}\hat k_2^{2n_2}}{(\hat k^2+\epsilon^2)^{p+\delta}}
      \quad\text{with}\quad
      \epsilon\ll 1,\; p\in\ZZ,\; n\in \ZZ_+^3\;,
      \label{eq:BInteg}
 \end{equation}
 can be reduced to a linear combination of two constants, \begin{equation}
      Z_0 = \cB_0(1;\{0,0,0\}) \approx 0.252731009858663\;
      \quad \text{and} \quad
      Z_1 =\frac{\cB_0(1;\{1,1,0\})}{3} \approx 0.181058342883210\;.
    \end{equation}
    Here, $\hat{k} = \frac{2}{a} \sin (ka/2)$ is the lattice momentum. These two constants have been determined to high precision using the Lüscher-Weisz coordinate-space method~\citep{Luscher:1995zz}.
   
The two momentum-dependent scalar lattice integrals required for the following LPT calculations are
    \begin{align}
      V(q) =&\int_{-\pi/a}^{\pi/a}\dk\frac{1}{(\hat k^2+m^2)(\wh{q-k}^2+m^2)}\,,
      \\[.5em]
      I_\mn(q) =&\int_{-\pi/a}^{\pi/a}\dk\frac{\bar k_\mu \ol{(q-k)}_\nu}{(\hat k^2+m^2)(\wh{q-k}^2+m^2)}\;,
    \end{align}
where $\bar{k} = \frac{1}{a} \sin (ka)$. By expanding the expressions in powers of the external momenta~\citep{Kawai:1980ja,Capitani:2002mp} and using the recursion relations, in the massless limit, these evaluate to
    \begin{equation} \label{eq:VIntSol}
     \lim_{m \to 0} V(q) = \frac{1}{8q}
      +a\pfrac{14Z_0+9Z_1-4}{96} 
      + a^3q^2\pfrac{34Z_0-9Z_1+4}{27648} + \ord{a^5q^4}
    \end{equation}
    \begin{multline}
    	\label{eq:IIntSol}
      \lim_{m \to 0}  I_\mn(q) = \frac{\delta_\mn}{a}\pfrac{1-6Z_0}{12}
      +\frac{q}{64}\pq*{2\delta_\mn-\pi_\mn}
      +a\bq*{\frac{Z_0}{16}\;\delta_\mn q_\mu^2 
      +q^2(2\pi_\mn-3\delta_\mn)\pfrac{10Z_0-9Z_1+4}{1152}}
      \\
      +\frac{a^3}{552960}\Big[
        4\pq*{\delta_\mn(q_1^4+q_2^4+q_3^4)+6q^2q_\mu^2\delta_\mn+4q_\mu q_\nu(q_\mu^2+q_\nu^2)}
        (52-86Z_0-117Z_1)
        \\+360\delta_\mn q_\mu^4(6Z_0+9Z_1-4)
        +q^4(4\pi_\mn-5\delta_\mn)(236-298Z_0-531Z_1)
      \Big]
      +\ord{a^5 q^6}.
    \end{multline}

\subsection{EMT operator mixing: $c_3$} \label{sec:app_emt_c3}

Here we calculate the perturbative renormalization of $\Tmn$ on the lattice. The na\"ive discretization of the EMT is
\begin{align}
\Tmn^0 = \frac{N}{g} \Tr \left\{ 2(\overbar{\delta}_\mu \phi)(\overbar{\delta}_\nu \phi) - \delta_\mn \left[ \sum_\rho (\overbar{\delta}_\rho \phi)^2 + (m^2 - m_c^2) \phi^2 + \phi^4 \right] + \xi \left( \delta_\mn \sum_\rho (\bar{\delta}_\rho \phi )^2 - (\bar{\delta}_\mu \phi)(\bar{\delta}_\nu \phi)\right) \right\}
\end{align}
Here the term multiplying $\xi$ is the improvement term. Since the improvement term is identically conserved and satisfies the Ward identities, $\xi$ has been taken to be 0 in the main text. The calculations here retrace the steps taken for the $4d$ case in~\citep{Caracciolo:1988hc}. 

\begin{figure}[h]
  \includegraphics[scale=1.0]{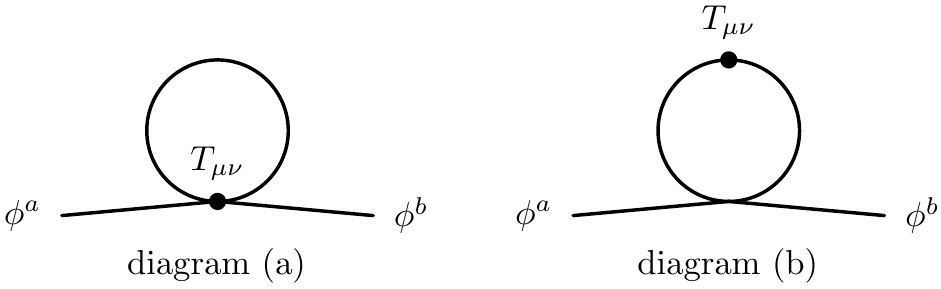}
  \caption{The insertion of $\Tmn$ in a two-point correlator, i.e. $\langle \phi^a(x_1) \phi^b(x_2) \Tmn(x_3) \rangle$, up to one loop. The black dot represents the insertion of $\Tmn$. }
  \label{fig:c3-1loop}
\end{figure}

By considering operators which have a lower dimension than $\Tmn$, the only operator capable of producing divergent mixing is $\delta_\mn \Tr \phi^2$. We therefore defined the renormalized EMT in~\cref{eq:renormalised-emt} as
\begin{align}
\TRmn &= \Tmn^0 - C_3 \delta_\mn \frac{N}{g} \Tr \phi^2,
\end{align}
with $C_3$ being the divergent mixing coefficient. To calculate the mixing coefficient perturbatively, consider the insertion of $\Tmn$ in the two-point correlator, i.e. $\langle \phi^a(x_1) \phi^b(x_2) \Tmn(x_3) \rangle$. The one-loop diagrams are shown in~\cref{fig:c3-1loop}. Both in the continuum and on the lattice, diagram (a) in~\cref{fig:c3-1loop} is finite, and contributes to the WI. However, for diagram (b), as a result of the breaking of translational invariance, the LPT result diverges, even though in the continuum the PT result is finite (this could be calculated by replacing the lattice momenta $\hat{q}$ with the continuum momenta $q$, and the integration limit by $\int_{-\infty}^\infty$).

Using LPT, diagram (b) in~\cref{fig:c3-1loop} evaluates to 
\begin{align} \label{eq:Bmn}
B_\mn(q) &= -\delta_{ab}\left( 2N- \frac{3}{N}\right) \left[ - 2I_\mn(q)+\delta_\mn\left(\sum_{\rho}I_{\rho\rho}(q)+m^2 V(q) \right) + \xi \left( I_\mn (q) - \delta_\mn \sum_\rho I_{\rho \rho}(q) \right) \right].
\end{align}

Using~\citep{Kawai:1980ja}, the divergent term of $B_\mn(q)$ can be isolated with $B_\mn(0)$, while the remaining terms are finite or vanish in the continuum limit. For $\xi=0$, this evaluates to
\begin{align} \label{eq:Bmn-div}
B_\mn(0)  &= \frac{\delta_\mn}{a}\left( 2N- \frac{3}{N}\right)\left( \frac{6Z_0-1}{12}\right) \nonumber \\
&= \frac{C_3^{\text{1 loop}}N}{g}\delta_\mn.
\end{align}
Using the definition from~\cref{eq:C3c3-1loop},
\begin{align}
C_3^{\text{1 loop}} = \frac{g}{a}  c_3^{\text{1 loop}},
\end{align}
to absorb the leading $\frac{1}{a}$ behavior, we obtain
  \begin{align}
c_3^{\text{1 loop}}&=\left(2-\frac{3}{N^2}\right)\left(\frac{6Z_0-1}{12}\right). 
\end{align}   
This gives the result in~\cref{eq:c3-1loop}.

\subsection{Correlators at vanishing flow time: $C_2(q)$ and $C_\mn(q)$} \label{sec:app_c2cmn_0t}
\begin{figure}[h]
\centering
\includegraphics[scale=1.0]{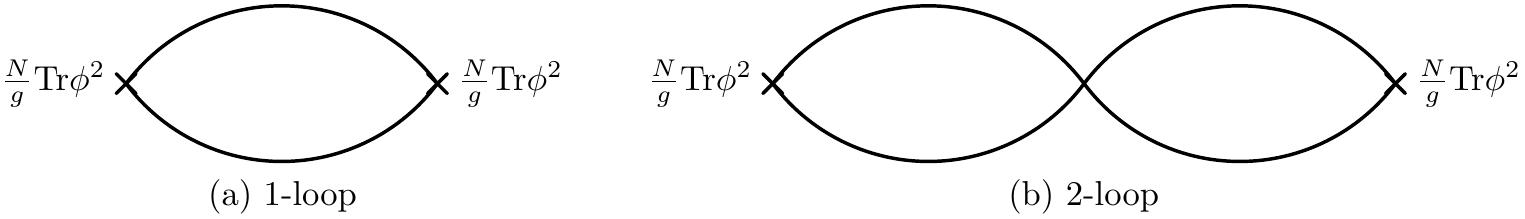}
\caption{Perturbative expansion of  $C_2(q)$ at one and two loops. }
\label{fig:trphi2trphi2}
\end{figure}

The first two-point correlation function to calculate is defined in~\cref{eq:C2}:
\begin{equation}
	C_2(q) =\left( \frac{N}{g} \right)^2 a^3 \sum_{x \in \Lambda} e^{-i q \cdot x} \langle \Tr \phi^2 (x) \Tr \phi^2 (0)\rangle. 	
\end{equation}
The one- and two-loop diagrams are shown in~\cref{fig:trphi2trphi2}(a) and (b) respectively.

Note that the two-loop diagram is simply the square of the one-loop diagram up to an overall color factor. These diagrams evaluate to 
\begin{align}
	C_2^\text{1 loop}(q)&=\tr(T_a T_b)\tr(T_c T_d)(\delta_{ac}\delta_{bd}+\delta_{ad}\delta_{bc})V(q)=\frac{N^2}{2}\pq*{1-\frac{1}{N^2}}V(q),\label{eq:C2basic}
	\\[.5em]
	C_2^\text{2 loop}(q)&= -2\pfrac{g}{N}\tr(T_a T_b)\tr(T_c T_d T_e T_f)\tr(T_g T_h)\left(\delta_{ac}(16\delta_{bd}\delta_{eg} + 8 \delta_{be}\delta_{dg})\delta_{fh}\right)V(q)^2
	\nonumber\\&=-N^2g\pq*{2-\frac{5}{N^2}+\frac{3}{N^4}}V(q)^2.
\end{align}
In the massless limit, using~\cref{eq:VIntSol}, these yield
\begin{align}
	C_2^\text{1 loop}(q) &= \frac{N^2}{16g}\left(1-\frac{1}{N^2}\right)\left[(g/q)+(ag)\frac{14Z_0+9Z_1-4}{12} + (ag)^3 (q/g)^2\frac{34Z_0-9Z_1+4}{3456}+\ord{(ag)^5}\right],
	\label{eq:C21lp}
	\\
	C_2^\text{2 loop}(q) &=-\frac{N^2}{64g}\left(2-\frac{5}{N^2}+\frac{3}{N^4}\right)\Big[
	(g/q)^2 
	+(ag)(g/q)\left(\frac{14Z_0+9Z_1-4}{6}\right)
	\nonumber\\&\hspace*{3cm}+
	(ag)^2\frac{(14Z_0+9Z_1-4)^2}{144}+
	(ag)^3(q/g)\left(\frac{34Z_0-9Z_1+4}{1728}\right)+\ord{(ag)^4}
	\Big].
	\label{eq:C22lp}
\end{align}

Now we evaluate the correlation function in~\cref{eq:Cmn}:
	\begin{align} \label{eq:appendix_Cmn}
	C_\mn(q)  = \frac{N}{g} a^3 \sum_{x \in \Lambda} e^{-i q \cdot x} \langle \Tmn^R (x) \Tr \phi^2 (0) \rangle = C_\mn^0(q) - \frac{gc_3}{a} \delta_\mn  C_2(q),
\end{align}
 where
\begin{align}
C_\mn^0(q) = \frac{N}{g} a^3 \sum_{x \in \Lambda} e^{-i q \cdot x} \langle \Tmn^0 (x) \Tr \phi^2 (0) \rangle,  \\
C_2(q) =\left( \frac{N}{g} \right)^2 a^3 \sum_{x \in \Lambda} e^{-i q \cdot x} \langle \Tr \phi^2 (x) \Tr \phi^2 (0)\rangle 
\end{align}

\begin{figure}[h]
\centering
\includegraphics[scale=1.0]{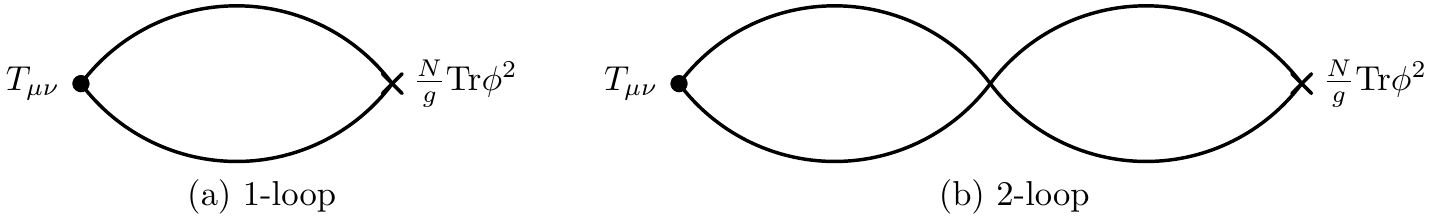}
\caption{Perturbative expansion of  $C_\mn^0 (q)$ at one and two loops. }
\label{fig:Tmntrphi2}
\end{figure} 
The relevant one- and two-loop diagrams for the correlator $C_\mn^0(q)$ are shown in~\cref{fig:Tmntrphi2}(a) and (b) respectively, and they evaluate to 
\begin{align}
	C^{0\text{ 1 loop}}_\mn(q) &= \frac{N^2}{2}\left(1-\frac{1}{N^2}\right)\Big[- 2I_\mn(q)+\delta_\mn\sum_{\rho}I_{\rho\rho}(q)+\delta_\mn m^2 C_2^\text{1 loop}(q)
	 	\nonumber\\&\hspace*{3cm}
	+ \xi \big( I_\mn (q) - \delta_\mn \sum_\rho I_{\rho \rho}(q)\big)  \Big]
	\label{eq:BareCmunu1lp}
	\\
	C^{0\text{ 2 loop}}_\mn(q) &= -N^2g\left(1-\frac{5}{2N^2}-\frac{3}{2N^4}\right)\Big\{\big[- 2I_\mn(q)+\delta_\mn\sum_{\rho}I_{\rho\rho}(q)\big]V(q)+\delta_\mn m^2 C_2^\text{2 loop}(q)
	 	\nonumber\\&\hspace*{3cm}
	+ \xi \big( I_\mn (q) - \delta_\mn \sum_\rho I_{\rho \rho}(q)\big)V(q)  \Big\}
	\,.
	\label{eq:BareCmunu2lp}
\end{align}
At one loop,~\cref{eq:BareCmunu1lp}, $C^0_\mn$ contains only the tree-level EMT, so $C_\mn^\text{1 loop}=C^{0\text{ 1 loop}}_\mn$. There is no contribution coming from the operator mixing $c_3$, which comes with another order $O(g)$. However, the term $\delta_\mn\sum_{\rho}I_{\rho\rho}(q) - 2I_\mn(q)$ presents a divergent contact term at $C^{0\text{ 1 loop}}_\mn(0)$,
\begin{align}
C_\mn^{0\text{ 1 loop}}(0) &= -\frac{N^2}{2a}\left(1-\frac{1}{N^2}\right)\left(\frac{6Z_0-1}{12}\right)\delta_\mn \nonumber \\
&= \frac{\kappa}{a} \delta_\mn.
\end{align}
The integral producing this contact term is similar to that in $c_3^{\text{1 loop}}$ in ~\cref{eq:Bmn-div}, with the only difference being the color factor. This contact term has to be subtracted before the continuum limit of the correlator is taken.

For the two-loop expression, it can be shown that after subtracting the correlator $\frac{gc_3}{a}\delta_\mn C_2^{\text{1 loop}}(q)$ to renormalize the EMT from ~\cref{eq:appendix_Cmn}, the correlator is UV finite; no extra divergences other than the one coming from the operator expansion appear.

\subsection{Correlators at finite flow time: $C_2(t, q)$, $C_\mn(t, q)$} \label{sec:app_c2cmn_ft}
At finite flow time, the lattice integrals are regulated by the flow time $t$. In perturbation theory, the kernel for each propagator has an extra exponential factor, $e^{-tq^2}$, where $q$ is the momentum of the propagator. We first evaluate the correlator  
\begin{align}
C_2(t, q) =\left( \frac{N}{g} \right)^2 a^3 \sum_{x \in \Lambda} e^{-i q \cdot x} \langle \Tr \phi^2 (x) \Tr \rho^2 (t, 0)\rangle
\end{align}
at finite flow time. This correlator is obtained by replacing $\Tr\phi^2(x)$ with $\Tr\rho^2(t,0)$ in $C_2(q)$. Since the regulated correlators are finite, we look at the continuum limit ($a \to 0$) of the correlator in perturbation theory. At one loop, this evaluates to
\begin{align} \label{eq:append-flow-c2}
	C_2^\text{1 loop}(t, q) = \frac{N^2}{16}\left(1-\frac{1}{N^2}\right) \int \dmom{k} \frac{e^{-tq^2}e^{-t(q-k)^2}}{( k^2+m^2)((q-k)^2+m^2)}.
\end{align}
In the massless limit,
\begin{align}
C_2^\text{1 loop}(t, q)	= \frac{N^2}{16g}\left(1-\frac{1}{N^2}\right)\left[1-\frac{2}{\pi}\int_0^\sigma\mathrm{d}s\;\frac{e^{-2s^2}\erfi(s)}{s}\right]\left(\frac{g}{q}\right),
\end{align}	
where $\sigma = \sqrt{t q^2/2}$, and $\erfi (z) = -i \erf (iz)$ is the \textit{imaginary error function}, which has the series expansion $\erfi (z) = \pi^{-1/2}\left(2z+ \frac{2}{3}z^3+\cdots\right)$ about $z=0$. Expanding in $\sigma$, this evaluates to
\begin{align}\label{eq:append-flow-c2-expansion}
C_2^\text{1 loop}(t, q)	\approx \frac{N^2}{16g}\left(1-\frac{1}{N^2}\right)\left[1+\left(\frac{32q^2 t}{\pi^3}\right)^{1/2} 
	\left(\frac{5q^2 t}{18} -1\right)\right]\left(\frac{g}{q}\right)+\ord{\sigma^{5}}.
\end{align}

Similarly, we look at the continuum limit of 
\begin{align}
C_\mn^0(t, q) = \frac{N}{g}a^3 \sum_{x \in \Lambda} e^{-i q \cdot x} \langle \Tmn^0 (x) \Tr \rho^2 (t, 0) \rangle
\end{align}
at finite flow time. In the continuum limit, the EMT does not require renormalization; we can therefore drop the $0$ superscript. At one loop, 
\begin{align}
		C_\mn^{\text{1 loop}} (t, q) =\frac{N^2}{2}\left(1 - \frac{1}{N^2}\right)\int\frac{d^3 k}{(2\pi)^3}
		\frac{
			\delta_\mn k\cdot(q-k) - 2 k_\mu (q-k)_\nu 
			+\xi(q_\mu q_\nu -\delta_\mn q^2)
		}{(k^2+m^2)((q-k)^2+m^2)}
		e^{-tq^2}e^{-t(q-k)^2}
	\end{align}
In the massless limit, this evaluates to 
	\begin{multline}
		C_\mn^{\text{1 loop}} (t, q) = -\frac{N^2}{2}\left(1 - \frac{1}{N^2}\right)
		\frac{q}{64\pi^{3/2}}\Bigg[
			\sqrt{\pi}\erfi(\sigma)(\pi_\mn(3+2\sigma^2) - 2\delta_\mn)\sigma^{-4}e^{-2\sigma^2}
			\\
			+8\sqrt{\pi}(1-4\xi)\pi_\mn\int_0^\sigma \frac{e^{-2s^2}\erfi(s)}{s}\mathrm{d}s
			\\
			-2(1-4\xi)\pi^{3/2}\pi_\mn
			+e^{-\sigma^2}(4\delta_\mn-6\pi_\mn)\sigma^{-3}
		\Bigg]
	\end{multline}
where $\pi_\mn = \delta_\mn - \frac{q_\mu q_\nu}{q^2}$ is the transverse projector. 
	To obtain the `flowed contact term' $K(t)$ from~\cref{eq:hat-Cmn}, we utilize the fact that the contact term is the longitudinal part of the correlator $C_\mn(t,q)$. We separate the above expression for $C_\mn(t,q)$ into a transverse part, $C_\mn(t, q)^{\text{transverse}}$ (which is proportional to the $\pi_\mn$), and the remaining longitudinal part $C_\mn(t, q)^{\text{longitudinal}}$. The transverse part
	\begin{align}
	C_\mn(t, q)^{\text{transverse}} = &-\pi_{\mn}\frac{N^2}{2}\left(1 - \frac{1}{N^2}\right)
		\frac{q}{64\pi^{3/2}}\Bigg[
			\sqrt{\pi}\erfi(\sigma)(3+2\sigma^2 )\sigma^{-4}e^{-2\sigma^2}
			\nonumber\\
			&+8\sqrt{\pi}(1-4\xi)\int_0^\sigma \frac{e^{-2s^2}\erfi(s)}{s}\mathrm{d}s
			\nonumber\\
			&-2(1-4\xi)\pi^{3/2}
			+6e^{-\sigma^2}\sigma^{-3}
		\Bigg] 
	\end{align}
	is finite, as ensured by the WI. The remaining longitudinal part gives
	\begin{align} \label{eq:Kt}
	C_\mn(t, q)^{\text{longitudinal}} = -\delta_\mn \frac{N^2}{2}\left(1 - \frac{1}{N^2}\right)
		\frac{q}{64\pi^{3/2}}\Bigg[
			\sqrt{\pi}\erfi(\sigma)(- 2)\sigma^{-4}e^{-2\sigma^2}
			+4e^{-\sigma^2}\sigma^{-3}
		\Bigg].
	\end{align}
When expanded about $\sigma=0$, the leading order term contributing to the contact term $C_\mn(t, q)^{\text{longitudinal}}$ is
\begin{align}
C_\mn(t, q)^{\text{longitudinal}} &\approx -\delta_\mn \frac{N^2}{2}\left(1 - \frac{1}{N^2}\right) \frac{q}{64\pi^{3/2}} \frac{8}{3\sigma} + \ord{\sigma}\nonumber\\
		&= -\delta_\mn \frac{N^2}{2}\left(1 - \frac{1}{N^2}\right) \frac{\sqrt{2}}{24\pi^{3/2}\sqrt{t}} + \ord{\sigma}\nonumber\\
		&= \delta_\mn K(t) \label{eq:flow-kappa},
\end{align}
which gives us the result in~\cref{eq:flowtime-omega}.

Using~\cref{eq:append-flow-c2-expansion,eq:Kt}, the perturbative expression for the ratio in~\cref{eq:fg-ratio} can be calculated,
\begin{align}
f_g(g\sqrt{t}, q_l) &= \frac{a}{g} \frac{K(t)}{C_2(t,q_l)}\nonumber \\
&= -\frac{\sqrt{2}}{3 \pi^{3/2}} \frac{aq_l}{g\sqrt{t}} + \ord{\sigma},
\end{align}
giving the result in~\cref{eq:flow-fg}.

  \bibliography{cosmhol-emt}
\end{document}